\documentclass[longauth]{aa}
\usepackage{CJKutf8}
\usepackage{hyperref}

\def\arcmin{\ifmmode {^{\prime}}\else $^{\prime}$\fi}
\def\arcsec{\ifmmode {^{\prime\prime} }\else $^{\prime\prime}$\fi}
\usepackage{graphicx}
\usepackage[normalem]{ulem}
\usepackage{txfonts}
\usepackage{placeins}
\usepackage{xcolor}
\usepackage{multirow}
\usepackage{pdflscape}
\usepackage{longtable}
\usepackage{booktabs}
\begin{document}

\begin{CJK*}{UTF8}{gbsn}

   \title{The RAdio Galaxy Environment Reference Survey (RAGERS) \thanks{Tables \ref{850table} and \ref{multi-z} are only available in electronic form at the CDS via anonymous ftp to \url{cdsarc.u-strasbg.fr} (\url{130.79.128.5}) or via \url{http://cdsweb.u-strasbg.fr/cgi-bin/qcat?J/A+A/.}}}

   \subtitle{Evidence of an anisotropic distribution of submillimeter galaxies \\in the 4C\,23.56 protocluster at $z=2.48$}

   \author{D.~Zhou
          \inst{1,2,3}
          \and
          T.~R.~Greve
          \inst{1,2,4}
          \and 
          B.~Gullberg
          \inst{1,2}
          \and 
          M.~M.~Lee
          \inst{1,2}
          \and
          L.~Di~Mascolo
          \inst{5}
          \and
          S.~R.~Dicker
          \inst{6}
          \and
          C.~E.~Romero
          \inst{6,7,8}
          \and
          S.~C.~Chapman
          \inst{3,9,10}
          \and
          C.-C.~Chen 
          \inst{11}
          \and
          T.~Cornish
          \inst{12,13}
          \and
          M.~J.~Devlin
          \inst{6}
          \and 
          L.~C.~Ho 
          \inst{14,15}
          \and
          K.~Kohno
          \inst{16,17}
          \and
          C.~D.~P.~Lagos
          \inst{18,19,1}
          \and
          B.~S.~Mason
          \inst{20}
          \and
          T.~Mroczkowski
          \inst{21}
          \and
          J.~F.~W.~Wagg
          \inst{22}
          \and
          Q.~D.~Wang
          \inst{23}
          \and
          R.~Wang
          \inst{14,15}
          \and
          M.~Brinch
          \inst{1,2}
          \and
          H.~Dannerbauer
          \inst{24,25}
          \and
          X.-J.~Jiang 
          \inst{26,27}
          \and
          L.~R.~B.~Lauritsen
          \inst{28}
          \and
          A.~P.~Vijayan
          \inst{1,2}
          \and
          D.~Vizgan
          \inst{29,1,2}
          \and
          J.~L.~Wardlow
          \inst{12}
          \and
          C.~L.~Sarazin
          \inst{30}
          \and
          K.~P.~Sarmiento
          \inst{6}
          \and
          S.~Serjeant
          \inst{28}
          \and
          T.~A.~Bhandarkar 
          \inst{6}
          \and
          S.~K.~Haridas 
          \inst{6}
          \and
          E.~Moravec 
          \inst{8}
          \and
          J.~Orlowski-Scherer 
          \inst{6}
          \and
          J.~L.~R.~Sievers 
          \inst{31}
          \and
          I.~Tanaka
          \inst{32}
          \and
          Y.-J.~Wang
          \inst{11,33}
          \and
          M.~Zeballos 
          \inst{34,35}
          \and
          A.~Laza-Ramos 
          \inst{24,25}
          \and
          Y.~Liu
          \inst{14,15}
          \and
          M.~S.~R.~Hassan
          \inst{36}
          \and
          A.~K.~M.~Jwel
          \inst{36}
          \and
          A.~A.~Nazri
          \inst{36}
          \and
          M.~K.~Lim
          \inst{36}
          \and
          U.~F.~S.~U.~Ibrahim 
          \inst{36}
          }

   \institute{
                Cosmic Dawn Center (DAWN)      
\and
   DTU Space, Technical University of Denmark, Elektrovej 327, DK-2800 Kgs. Lyngby, Denmark 
  \\ \email{dzhou.astro@gmail.com}
  \and
             Department of Physics and Astronomy, University of British Columbia, 6225 Agricultural Rd., Vancouver, V6T 1Z1, Canada
             \and
            Department of Physics and Astronomy, University College London, Gower Street, London WC1E 6BT, United Kingdom
             \and
             Laboratoire Lagrange, Université Côte d’Azur, Observatoire de la Côte d’Azur, CNRS, Blvd de l’Observatoire, CS 34229, 06304 Nice cedex 4, France
             \and
             Department of Physics and Astronomy, University of Pennsylvania, 209 South 33rd Street, Philadelphia, PA, 19104, USA
             \and
             Center for Astrophysics | Harvard and Smithsonian, 60 Garden Street, Cambridge, MA, 02143, USA
             \and
             Green Bank Observatory, 155 Observatory Road, Green Bank, WV 24944, USA
             \and
             National Research Council, Herzberg Astronomy and Astrophysics, 5071 West Saanich Rd., Victoria, V9E 2E7, Canada
             \and
             Department of Physics and Atmospheric Science, Dalhousie University, Halifax, Nova Scotia, B3H 4R2, Canada
             \and
             Academia Sinica Institute of Astronomy and Astrophysics (ASIAA), No. 1, Section 4, Roosevelt Road, Taipei 10617, Taiwan
             \and
             Department of Physics, Lancaster University, Lancaster, UK
             \and
             Department of Physics, University of Oxford, Denys Wilkinson Building, Keble Road, Oxford OX1 3RH, UK
             \and
             Department of Astronomy, School of Physics, Peking University, Beijing 100871, People's Republic of China
             \and
             Kavli Institute for Astronomy and Astrophysics, Peking University, Beijing 100871, People's Republic of China
             \and
             Institute of Astronomy, Graduate School of Science, The University of Tokyo, 2-21-1 Osawa, Mitaka, Tokyo 181-0015, Japan
             \and
             Research Center for the Early Universe, Graduate School of Science, The University of Tokyo, 7-3-1 Hongo, Bunkyo-ku, Tokyo 113-0033, Japan
             \and
             International Centre for Radio Astronomy Research, The University of Western Australia, 35 Stirling Hwy, 6009, Crawley, WA, Australia
             \and
             ARC Centre of Excellence for All Sky Astrophysics in 3 Dimensions (ASTRO 3D), Australia
             \and
             National Radio Astronomy Observatory, 520 Edgemont Rd., Charlottesville, VA 22903, USA
             \and
             European Southern Observatory, Karl-Schwarzschild-Strasse 2, Garching D-85748, Germany
             \and
             PIFI Visiting Scientist, Purple Mountain Observatory, Np. 8 Yuanhua Road, Qixia District, Nanjing 210034, People's Republic of China
             \and
             Department of Astronomy, University of Massachusetts, Amherst, MA 01003, USA
             \and
             Instituto de Astrofísica de Canarias (IAC), E-38205 La Laguna, Tenerife, Spain
             \and
             Departamento Astrofísica, Universidad de la Laguna, E-38206 La Laguna, Tenerife, Spain
             \and
             Research Center for Intelligent Computing Platforms, Zhejiang Laboratory, Hangzhou 311100, People's Republic of China
             \and
             East Asian Observatory, 660 North A'ohoku Place, Hilo, Hawaii, 96720, USA
             \and
             School of Physical Sciences, The Open University, Walton Hall, Milton Keynes MK7 6AA, UK
             \and
             Department of Astronomy, University of Illinois, 1002 West Green Street, Urbana, IL 61801, USA
             \and
             Department of Astronomy, University of Virginia, 530 McCormick Road, Charlottesville, VA 22904-4325, USA
             \and
             Department of Physics, McGill University, 3600 University Street, Montreal, QC, H3A 2T8, Canada
             \and
             Subaru Telescope, National Astronomical Observatory of Japan, 650 North A'ohoku Place, Hilo, HI 96720, USA
             \and
             Graduate Institute of Astrophysics, National Taiwan University, Taipei, Taiwan
             \and
             Instituto Nacional de Astrofísica Óptica y Electrónica, Luis Enrique Erro 1, Tonantzintla CP 72840, Puebla, México
             \and
             Universidad de las Américas Puebla. Ex Hacienda Sta. Catarina Mártir S/N. San Andrés Cholula, Puebla 72810, México
             \and
             Department of Physics, Universiti Malaya, 50603 Kuala Lumpur, Malaysia
             }
   \date{Received 6 November 2023 / Accepted 18 July 2024}
  \abstract
   {High-redshift radio(-loud) galaxies (H$z$RGs) are massive galaxies with powerful radio-loud active galactic nuclei (AGNs) and serve as beacons for protocluster identification. However, the interplay between H$z$RGs and the large-scale environment remains unclear.}
   {To understand the connection between H$z$RGs and the surrounding obscured star formation, we investigated the overdensity and spatial distribution of submillimeter-bright galaxies (SMGs) in the field of 4C\,23.56, a well-known H$z$RG at $z=2.48$.}
   {We used SCUBA-2 data ($\sigma\,{\sim}\,0.6$\,mJy) to estimate the $850\,{\rm \mu m}$ source number counts and examine the radial and azimuthal overdensities of the $850\,{\rm \mu m}$ sources in the vicinity of the H$z$RG.
   }
   {The angular distribution of SMGs is inhomogeneous around the H$z$RG 4C\,23.56, with fewer sources oriented along the radio jet.
   We also find a significant overdensity of bright SMGs (${\rm S}_{850\rm\,\mu m}\geq5\,$mJy). Faint and bright SMGs exhibit different spatial distributions. The former are concentrated in the core region, while the latter prefer the outskirts of the H$z$RG field. High-resolution observations show that the seven brightest SMGs in our sample are intrinsically bright, suggesting that the overdensity of bright SMGs is less likely due to the source multiplicity.
   }
   {}

   \maketitle
%

\section{Introduction}
Protoclusters are the high-redshift progenitors of galaxy clusters seen at the present day. These structures are thought to reside within massive dark matter halos and undergo violent relaxation during virialization \citep{Kravtsov2012}. These regions are believed to be where cluster galaxies formed, evolved, and quenched earlier than field galaxies \citep[e.g.,][]{Overzier2016}.

One of the classical techniques for finding protoclusters is to measure the overdensities of galaxies around high-redshift radio(-loud) galaxies \citep[H$z$RGs;][]{miley2008}. H$z$RGs are often embedded in massive dark matter halos and are thought to be the precursors of the brightest cluster galaxies in galaxy clusters \citep{Fanidakis2013,Hatch2014}.
Numerous protocluster campaigns searching for protoclusters using H$z$RGs as beacons have successfully found overdensities in a variety of galaxy populations, including extremely red objects, distant red galaxies, Lyman break galaxies, Ly$\alpha$ emitters, and H$\alpha$ emitters \citep[e.g.,][]{Venemans2007,Kodama2007, Hatch2011,Mayo2012,Wylezalek2013,Kotyla2016,Noirot2016,Noirot2018,Castignani2019,Uchiyama2022,Cordun2023}.

Galaxy number counts and clustering analyses suggest that radio-loud active galactic nuclei (AGNs) are likely to inhabit denser environments at cosmic noon  \citep[$z\,{\sim}\,1\,{-}\,3.5$; e.g.,][]{Hickox2009,Donoso2010,Hatch2014,Malavasi2015}. However, due to the potential selection bias toward the most extreme sources, this connection has not been confirmed \citep[e.g.,][]{Delvecchio2017,Thomas2022}, meaning it is unclear if the radio activity requires certain large-scale environments \citep[e.g.,][]{west1994,Hatch2014,Codis2018} or if it is an outcome of the AGN duty cycle in every massive galaxy \citep[e.g.,][]{lovell2018,Thomas2021,Delvecchio2022}.

The phenomenon of angular clustering around H$z$RGs is another open question. The surrounding galaxies have been observed to exhibit asymmetric distributions and show either angular concentration or avoidance in relation to the direction of the radio jet, which implies a connection between radio AGNs and large-scale structures \citep[e.g.,][]{Rees1989,west1994,Zeballos2018,Tozzi2022}. Such an angular nonuniformity has been found in both the local and high-redshift Universe \citep[e.g.,][]{Stevens2003,MartinNavarro2021,Stott2022,Uchiyama2022,Makoto2023,AB2023}. It is unclear, however, if it is caused by the powerful AGN feedback \citep[e.g.,][]{Kauffmann2015,Sorini2022} or because the radio jet orientation depends on the structure of the cosmic filaments \citep[e.g.,][]{west1994,Codis2018}. Therefore, studying the environments of radio galaxies is essential to understanding the interplay between the AGN activity and the environment.

One of the galaxy populations hosted by H$z$RG environments are submillimeter-bright galaxies \citep[SMGs; e.g.,][]{Smail1997, Blain2004,Chapman2004}, which are dust-obscured galaxies with intense star formation rates
\citep[$\gtrsim$\,100--1000\,M$_\odot$/year,][]{casey2014review}
and are luminous at far-infrared (FIR) wavelengths \citep[$S_{\rm 850\,\mu m}\,{\gtrsim}\,1\rm\,mJy$,][]{Hodge2020}.
They are believed to be the main contributors to the star formation activity in protoclusters and the ancestors of the massive elliptical galaxies seen in galaxy clusters today \citep{Ivison2000, Stevens2003}. 
Compared to other protocluster surveys, searches for SMG overdensities in H$z$RG environments have so far been only moderately successful \citep[e.g.,][]{Rigby2014,Zeballos2018}, largely due to the coarse angular resolution of single-dish submillimeter telescopes and the difficulty in obtaining redshifts of submillimeter sources.
The lack of easily accessible redshifts means that the sky-projected overdensities observed at submillimeter wavelengths are likely to be contaminated by foreground or background interlopers \citep[e.g.,][]{Chapman2015,Meyer2022}.

To construct a comprehensive understanding of cluster formation, we need to explore the interplay between SMGs and H$z$RGs, and their corresponding roles in protoclusters at different epochs. In particular, it is crucial to study such relations at cosmic noon ($z\,{\sim}\,1\,{-}\,3.5$), when both star formation and AGN activities reach their peaks and structures start collapsing \citep{Overzier2016}. 
The SCUBA-2 bolometer \citep[Submillimetre Common-User Bolometer Array 2,][]{holland2013} is a workhorse in such studies, thanks to its capability to map the overdensities of SMGs in megaparsec-scale environments (${\sim}\,10$\,cMpc), where most of the protocluster members are found \citep{chiang2013}.

The redshift is key to identifying protocluster members. Unfortunately, spectroscopic confirmation is observationally expensive and line mapping is less efficient for large sky surveys. However, with existing panchromatic data for multiple H$z$RG fields, it is feasible to estimate the photometric redshift (photo-$z$) and weed out a part of the SMGs not associated with protoclusters. This strategy facilitates the spectroscopic redshift confirmations for most protocluster SMGs within a reasonable observation time. To demonstrate the feasibility of this strategy, in this pilot study we utilized archival SCUBA-2 observations with multiwavelength ancillary data  as part of the RAdio Galaxy Environment Reference Survey (RAGERS). RAGERS is a large program (Program ID M20AL015) being carried out with the \textit{James Clerk} Maxwell Telescope (JCMT); it uses SCUBA-2 to target 27 powerful H$z$RGs uniformly distributed across the cosmic noon epoch. The RAGERS survey aims to significantly expand the current sample of submillimeter observations of H$z$RG fields in megaparsec-scale environments, with the goal of constraining the cosmic evolution of obscured star formation around H$z$RGs.

\begin{figure}[t]
    \centering
    \includegraphics[width=0.48\textwidth]{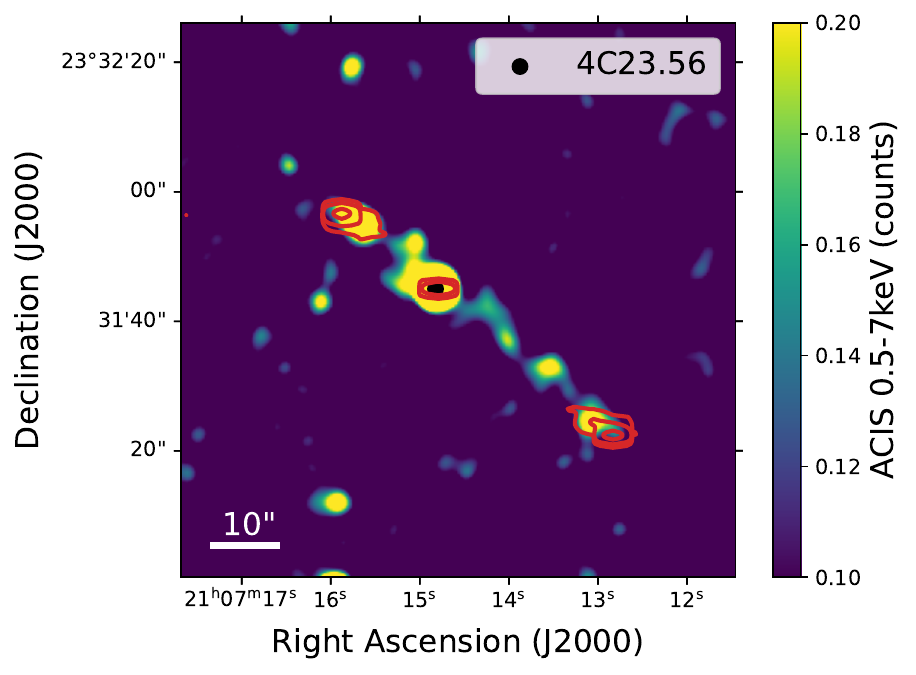}
    \caption{VLA (Very Large Array) 4.8~GHz contours (B configuration) of 4C\,23.56 overlaid on 0.5-7~keV \textit{Chandra} data, which both align in the northeast-southwest direction. The contour levels mark 4.8~GHz flux densities of 0.0001, 0.001, and 0.01\,Jy.}
    \label{fig:rj}
\end{figure}

In this work we investigated $850\,{\rm \mu m}$-selected SMGs in the well-known protocluster field 4C\,23.56 at $z\,{=}\,2.48$.
4C\,23.56 is a powerful Fanaroff-Riley II H$z$RG with extended X-ray and radio emissions (see Fig.~\ref{fig:rj}), where X-ray is mainly produced by inverse scattering of the cosmic microwave background photons, which indicates its prolonged and extensive AGN feedback \citep[e.g.,][]{Johnson2007,Blundell2011}.
Significant overdensities have been found in its surrounding environment \citep[e.g.,][]{Knopp1997,Kajisawa2006,Tanaka2011,Galametz2012,Mayo2012,Wylezalek2013}, further confirming the presence of a protocluster \citep{Lee2019}.

The paper is structured as follows. In Sect. \ref{s2} we describe the SCUBA-2 $850\,{\rm \mu m}$ data and the data processing strategy employed in this study. We present the ancillary data in Sect. \ref{sec3}. In Sect. \ref{s4} we outline our analysis methods and present our main findings. In Sect. \ref{s5} we discuss the cause of the anisotropic distribution, the potential origin of the excess bright SMGs. We present a summary of this study in Sect. \ref{s6}.
Throughout this paper, we use a flat $\Lambda$ cold dark matter cosmology \citep[$H_0\,{=}\,69.3\,{\rm km\,s^{-1}\,Mpc^{-1}}$, $\Omega_m\,{=}\,0.287$;][]{Hinshaw2013}.
\section{SCUBA-2 observations} \label{s2}
In this section we describe the strategies adopted for constructing the final SCUBA-2 $850\,{\rm \mu m}$ map and source catalog. To ensure a fair comparison with the blank field, we followed a similar approach as described by \citet{Geach2017}, employing the same detection threshold (3.5$\sigma$). We refer the reader to \citet{Geach2017} for a detailed description.

\subsection{Observations and data reduction}
The SCUBA-2 $850\,{\rm \mu m}$ observations for this study were conducted between August 22, 2012, and June 3, 2016 (Program IDs JCMT-LR, M12BU39, M15AI146, and M15BI053), under good weather conditions (${\rm \tau_{\rm 225GHz}}\,{<}\,0.07$). The total exposure times for the "Pong900" and "CV Daisy" scan patterns are $\sim$\,7\,hours and $\sim$\,10\,hours, respectively.

\begin{figure*}[!t]
    \centering
    \includegraphics[width=0.95\textwidth]{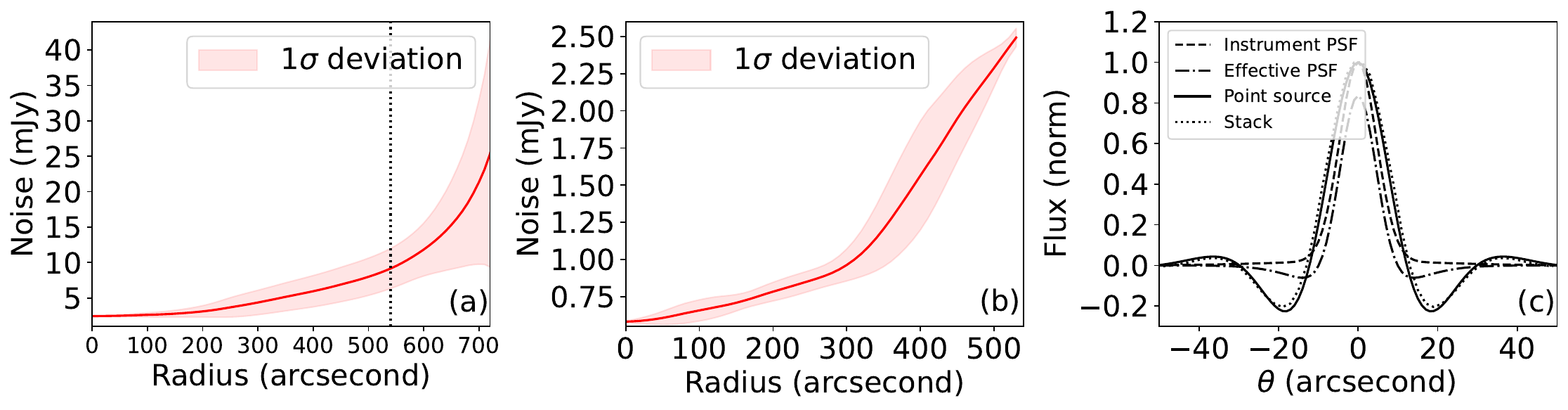}
    \caption{Overview of the noise and PSF properties before and after applying the matched filter. (a) Azimuthally averaged radial RMS profile of the SCUBA-2 map before the matched filtering process, with the 1$\sigma$ deviation shown as the shaded region. The RMS increases monolithically with radius. The dashed line reflects the radius of the retained region that we consider.
    {\bf (b)} RMS profile of the matched filtered map, which is significantly reduced compared to the original RMS level. The inhomogeneous radial RMS changes the effective area observed at different flux levels.
    {\bf (c)} Two-component instrumental PSF from \citet[dashed line]{Mairs2021}. The effective PSF for the matched-filtering process is shown as a dash-dotted line. The solid line and dotted line represent the shape of the point source from analytics and from stacking all sources above $5\,\sigma$ in the matched-filtered map, respectively. The broadened PSF and ``negative ring'' are caused by the smoothing process and background subtraction. }
    \label{fig:850-noise-profile-psf}
\end{figure*}
\begin{figure*}[!t]
    \centering
    \includegraphics[width=0.75\textwidth]{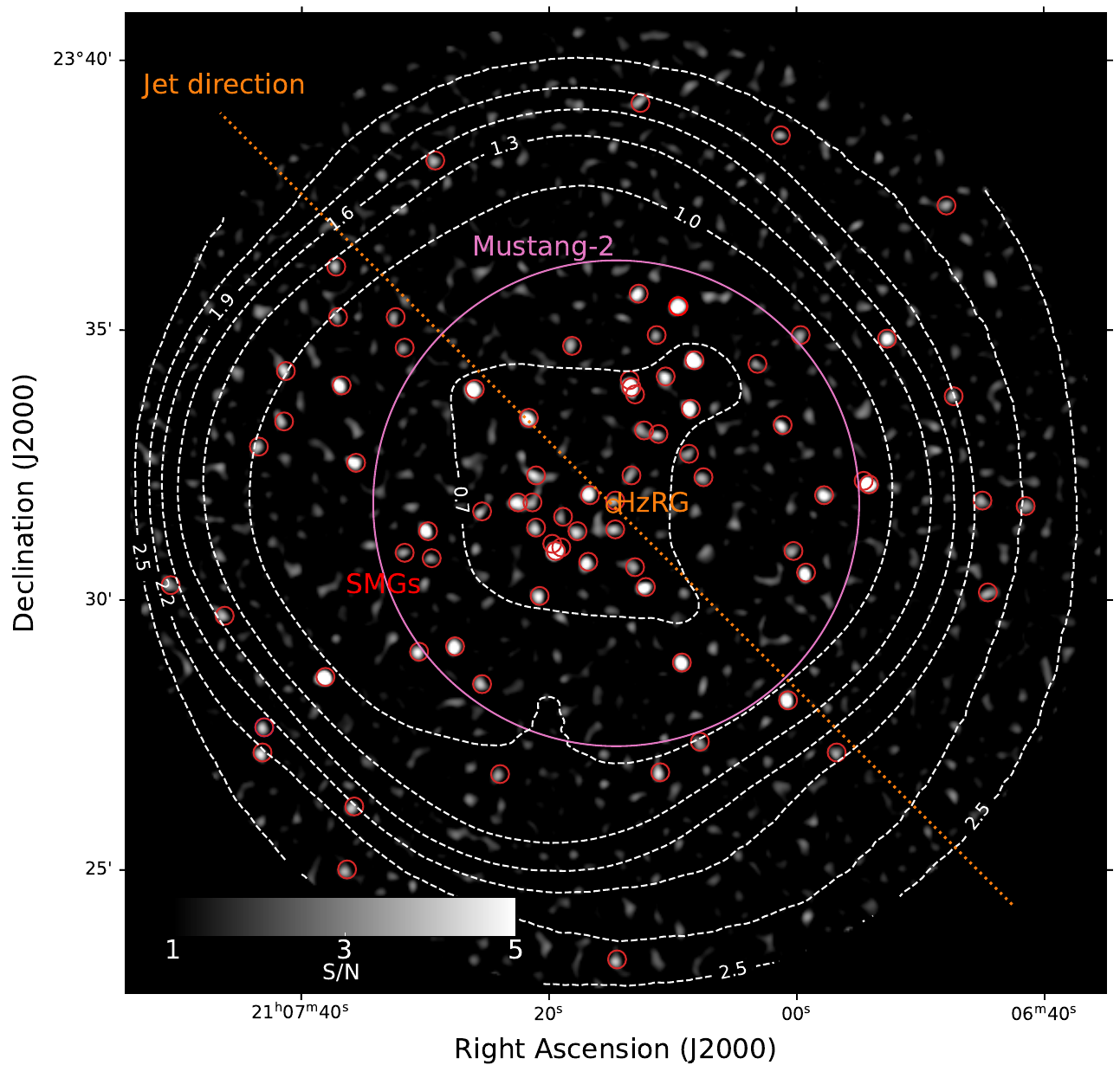}
    \caption{SCUBA-2 $850\,{\rm \mu m}$ S/N map of 4C\,23.56 ($z\,{=}\,2.48$) shown in grayscale. The solid pink contour shows the footprint of MUSTANG-2 observations.
    The noise contours are indicated as dashed white lines in units of mJy/beam. The H$z$RG and SMGs are marked as an orange circle and red circles, respectively. The corresponding jet direction (Fig.~\ref{fig:rj}) is shown as a dotted orange line.}
    \label{fig:ovlay}
\end{figure*}

We reduced the data using the standard pipeline of the STARLINK software package \citep[2021A;][]{Currie2014,Berry2022}. Figure\,\ref{fig:850-noise-profile-psf}a shows the azimuthally averaged radial root mean square (RMS) profile of the raw signal map. As part of the source extraction, the map was matched-filtered to enhance the point source detectability. Before the matched-filtering process, we first modeled the background and subtracted it from the SCUBA-2 map (see Sect. \ref{subsec:background-removal-mf} for details). As the noise level is significantly elevated at radii $\gtrsim9'$ from the map center, to ensure the effectiveness of the matched-filtering,
we cropped the map to retain the region within $9\arcmin$ of the center of the map (indicated by the dashed vertical line in Fig.\,\ref{fig:850-noise-profile-psf}a). The matched-filtering process significantly increases the sensitivity (Fig.\,\ref{fig:850-noise-profile-psf}b), but also broadens the shape of the point sources and results in a negative ring (see Fig.\,\ref{fig:850-noise-profile-psf}c).
The final central RMS noise level is $0.6\,{\rm mJy}$ and increases away from the center, reaching $2.5\,{\rm mJy}$ at the edge (Fig.\,\ref{fig:850-noise-profile-psf}b).
The SCUBA-2 $850\,{\rm \mu m}$ signal-to-noise ratio (S/N) map is presented in Fig.~\ref{fig:ovlay}.

\subsection{Background subtraction and matched-filtering} \label{subsec:background-removal-mf}
We subtracted the background by employing the Mexican hat wavelet technique \citep{Barnard2004,gonzalez2006}, which is identical to the matched-filter recipe in PICARD \citep{Gibb2013}. This removes the pattern noise and optimizes the point source extraction.
We first smoothed the original map by a large Gaussian kernel to estimate the background. Subsequently, we subtracted this smoothed map from the original signal map. The same procedure was applied to the corresponding point spread function (PSF) to estimate the effective PSF of the subtracted map. The matched-filtering and background subtraction broaden the effective PSF and cause a ``negative ring'' (see Fig.\,\ref{fig:850-noise-profile-psf}c).

\begin{figure}[t]
    \centering
    \includegraphics[width = 0.47 \textwidth]{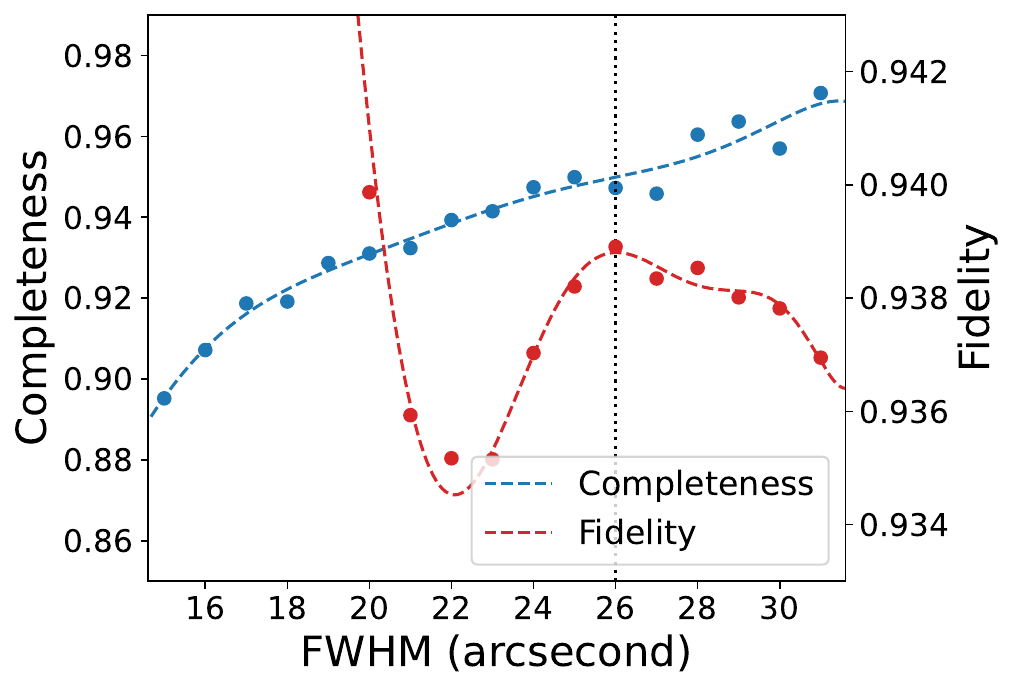}
    \caption{Overall completeness and fidelity as a function of Gaussian kernel size. The scatters are measured values, and the dashed lines represent the corresponding fitting results. To limit the potential contamination from spurious sources, we chose ${\rm FWHM}=26\arcsec$ as the optimal kernel size for background subtraction, which is indicated as the vertical dotted line.
    }
    \label{fig:BK850}
\end{figure}

However, the size of the kernel used for the background subtraction can affect the number of detected sources. We therefore used 10000 mock maps (see Sect. \ref{simulation})
with different sizes of Gaussian kernels to determine the optimal kernel size. We examined the completeness and fidelity (see Sect. \ref{comp_f}) as a function of kernel size. As shown in Fig.~\ref{fig:BK850}, completeness increases monotonically but fidelity reaches its local maximum at a full width at half maximum (FWHM) of $26\arcsec$. Consequently, we adopted $26\arcsec$ as the size of the large kernel to maintain a high completeness and avoid the excess of spurious sources.

Given the large beam size of JCMT \citep[${\rm FWHM}=12.6\arcsec$ at $850\,{\rm \mu m}$;][]{Mairs2021}, it is reasonable to treat SMGs as point sources in SCUBA-2 map.
To optimize the point source detection and flux measurements, we applied a matched-filtering technique, which enhances the point source detectability \citep{Serjeant2003}. In short, this technique returns the best-fit flux with minimized $\chi^2$.
The best-fit flux is given by
\begin{equation}
F=\frac{(S W) \otimes P}{W \otimes P^{2}},
\end{equation}
where $S$ is the original signal map; $P$ is the PSF of the instrument; and $W$ is the inverse-variance weight map; $\otimes$ denotes convolution. The corresponding flux error is calculated as
\begin{equation}
\Delta F=\frac{1}{\sqrt{W \otimes P^{2}}}.
\end{equation}
$F/\Delta F$ has also been calculated as the S/N value for source extraction.

\subsection{Source extraction} \label{se}
To extract source fluxes from the matched-filtered map, we applied a top-down algorithm similar to the one used by \citet{Geach2017}. This algorithm first identifies the maximum value in the matched-filtered S/N ($F/\Delta F$) map generated during the matched filtering process. If the value exceeds our selection threshold, we
measured the corresponding pixel value in the matched-filtered signal ($F$) map and recorded the coordinate, flux, S/N, and RMS noise of the source on our source catalog. Based on the S/N and flux value, we scaled the point source profile and subtracted it from the S/N and signal maps. This process iterates until the maximum value in the subtracted S/N map falls below 3.5\,$\sigma$, which is the standard detection threshold used in the SCUBA-2 blank field survey \citep[e.g.,][]{Geach2017, Simpson2019}. This method is effective for resolving blended sources. However, it is important to acknowledge the potential limitation: it
may incorrectly classify a mildly extended source as multiple point sources \citep[e.g.,][]{Wang2017}.
Additionally, we extracted sources with $3.5\,{\leq}\,{\rm S/N}\,{<}\,4$ for statistical purposes
but excluded them from catalog and further analysis due to the increasing fraction of spurious sources for ${\rm S/N}\,{<}\,4$ (see Sect.~\ref{debo}).

We note that \citet{Geach2017} report a ${\sim}\,10$\% flux loss introduced by the filtering step in PICARD recipe {\tt scuba2\_matched\_filter}. To assess this potential effect,
we injected a single artificial source at the center of a mock noise map, with the lowest noise level, to eliminate the noise contribution. We then applied our recipe and used our source extraction algorithm to obtain the source flux. After repeating this process 1,000 times for each flux level, we found the average recovered fraction beyond 0.97, even for the faintest source.
The level of flux loss is much less than the calibration uncertainty (${\sim}\,10\%$), supporting the reliability of flux measurement through our filtering and extraction procedure.

\subsection{Flux de-boosting, completeness, fidelity, and positional uncertainty} \label{debo}
We assessed the boosting factor and estimated the fidelity, completeness rate, and positional uncertainty through Monte Carlo simulations.
Due to the small number of sources in our field, which leads to significant Poisson error, we did not attempt to derive a new parameterized expression of number counts in a Schechter form or reevaluate the aforementioned parameters.
\subsubsection{Simulations} \label{simulation}
For generating the noise in our simulated maps, we assumed a Gaussian noise distribution and created maps following the RMS level of the science map. We produced several jackknife maps to confirm that the noise distribution closely matches a Gaussian distribution. We then inserted artificial sources at random positions in the simulated noise map, and adopted the number counts from \citet{Geach2017} as the flux density distribution in the blank field:
\begin{equation}
\frac{\mathrm{d} N}{\mathrm{~d} S}=\left(\frac{N_0}{S_0}\right)\left(\frac{S}{S_0}\right)^{-\gamma} \exp \left(-\frac{S}{S_0}\right),
\end{equation}
where $\mathrm{d} N/\mathrm{~d} S$ is the differential number counts, $S$ represents the flux density at $850\,{\rm \mu}$m, $N_0\,{=}\,7180\,{\rm deg}^{-2}, S_0\,{=}\,2.5\,{\rm mJy}$ and $\gamma\,{=}\,1.5$.

To determine the faintest flux for source injection in our simulation, we considered the minimum flux contributed to the extragalactic background light. \citet{Fujimoto2015} find that the main contributors at 1.2\,mm should be the sources above 0.02 mJy, which corresponds to ${\sim}\,0.05\,{\rm mJy}$ at $850\,{\rm \mu m}$. Hence, we injected sources with the minimum threshold of 0.05\,mJy and record their input fluxes, RMS, and positions.

Studies suggest that overdensities usually become negligible for faint sources in protocluster fields \citep[e.g.,][]{Lacaille2018, Garcia2020, Wang2021}. Assuming a homogeneous overdensity may be not accurate and could cause a potential bias.
Therefore, we injected faint sources ($S_{\rm 850\mu m}\,{<}\,5\,{\rm mJy}$) following the blank field number counts and have twice and three times the number of bright sources ($S_{\rm 850\mu m}\,{\geq}\,5\,{\rm mJy}$) in the outskirts ($r\,{\geq}\,4\arcmin$) and the central region ($r\,{<}\,4\arcmin$), which is also consistent with our results shown in Sect.\,\ref{ovd}.
\subsubsection{Flux de-boosting} \label{sec:deboosting}
The observed source flux can be boosted by noise spikes or fainter sources in the map, which is known as flux boosting \citep{Eddington1913}. To account for this, we investigated the ratio between input flux and recovered flux ($S_{\rm inp}/S_{\rm rec}$, i.e., the de-boosting factor) in our mock maps in flux bins of 1\,mJy and RMS bins of 0.1\,mJy.
We extracted sources from the matched-filtered mock maps and cataloged them in a manner consistent with our source extraction algorithm. These extracted sources were then matched with the input sources within a search radius of 6\arcsec, which is based on the average positional uncertainty \citep{Geach2017},
\begin{equation}
\delta \theta=1.2" \times\left(\frac{\mathrm{S/N}}{5}\right)^{-1.6}.
\label{eq:positional-offset}
\end{equation}
For sources just above our detection threshold (i.e., ${\rm S/N}\,{=}\,3.5$), the average positional uncertainty is ${\sim}\,2\arcsec$. A matching distance of 6\arcsec\ (${\sim}\,3\sigma$) is thus reasonable, as it does not result in a loss of many matched sources (${\lesssim}\,0.02\%$).

The matched sources are cataloged with the corresponding positional offsets.
We repeated this process 10,000 times to evaluate the boosting factor, and
find that the de-boosting factor follows a Gaussian distribution.
We therefore used the average value and standard deviation
as the de-boosting factor and associated uncertainty.

We obtained the de-boosting factor as a function of RMS noise and the observed flux density. For visualization, we show the de-boosting factor (solid red line) and its associated uncertainty (shaded region) as a function of the S/N (Fig.~\ref{fig:de}). We can see that boosting contributes less as S/N increases. In our source catalog (Table\,\ref{850table}), we have corrected the observed flux using the mean value of de-boosting factor at given flux and rms level and provide the uncertainty based on the $1\sigma$ value. To maintain statistical significance for subsequent number count calculations, we did not exclude any sources with de-boosted fluxes below $3.5\sigma$.

\subsubsection{Sample completeness and fidelity} \label{comp_f}
\begin{figure}[!t]
    \centering
    \includegraphics[width=0.47\textwidth]{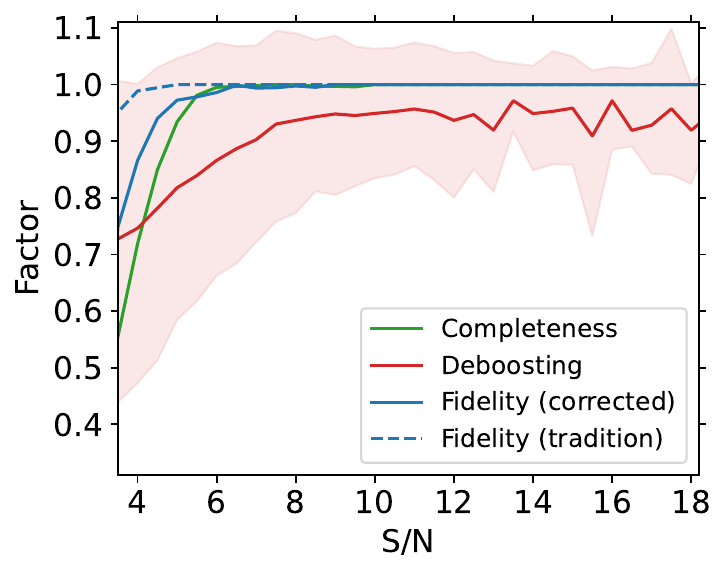}
    \caption{Estimations of completeness, de-boosting (1$\sigma$ uncertainty shown with the shaded region), and fidelity as a function of the S/N for SCUBA-2 sources in the 4C\,23.56 field. For comparison, the fidelity obtained by the traditional method is indicated as the dashed blue line, which is significantly higher than the updated value.}
    \label{fig:de}
\end{figure}

To assess the reliability of the source counts, we estimated the number of input sources $N_{\rm inp}$, recovered sources $N_{\rm rec}$, and spurious sources $N_{\rm spu}$ and then derived the completeness ($C\,{=}\,N_{\rm rec}/N_{\rm inp}$) and the fidelity ($f_{\rm fid}\,{=}\,1\,{-}\,N_{\rm spu}/N_{\rm rec}$) as functions of the input (recovered) flux and RMS level from the 10,000 mock maps. We calculated the values in each input (recovered) flux bin and RMS bin in each map and took the average as the completeness (fidelity) factor. The source counts are highly reliable for the significant detection (${>}\,5\sigma$; see Fig.~\ref{fig:de} for a visualization).

The number of spurious sources can be determined directly from the matched-filtered noise map, or by counting the recovered sources that do not match any input sources in our mock maps. In this work we adopted the latter approach. This is because values estimated from the matched-filtered noise maps can underestimate the contamination rate by a factor of 3 \citep[known as the multiple hypotheses problem; e.g.,][]{Vio2016,Vio2017,Vio2019}. The matched filtering technique increases the number of spurious sources in the signal map compared to the noise map because of the presence of peaks from astronomical signals. To illustrate this effect, we identified all recovered sources that do not match any injected sources ($>$1$\sigma$) as spurious sources when estimating the corrected fidelity (solid blue line in Fig.~\ref{fig:de}). This approach is similar to that of \citet{Casey2013}, though they use a $3\sigma$ detection limit. This adjustment results in a significantly lower fidelity compared to the traditional method.

\subsubsection{Positional uncertainty}
\begin{figure}[!t]
    \centering
    \includegraphics[width=0.47\textwidth]{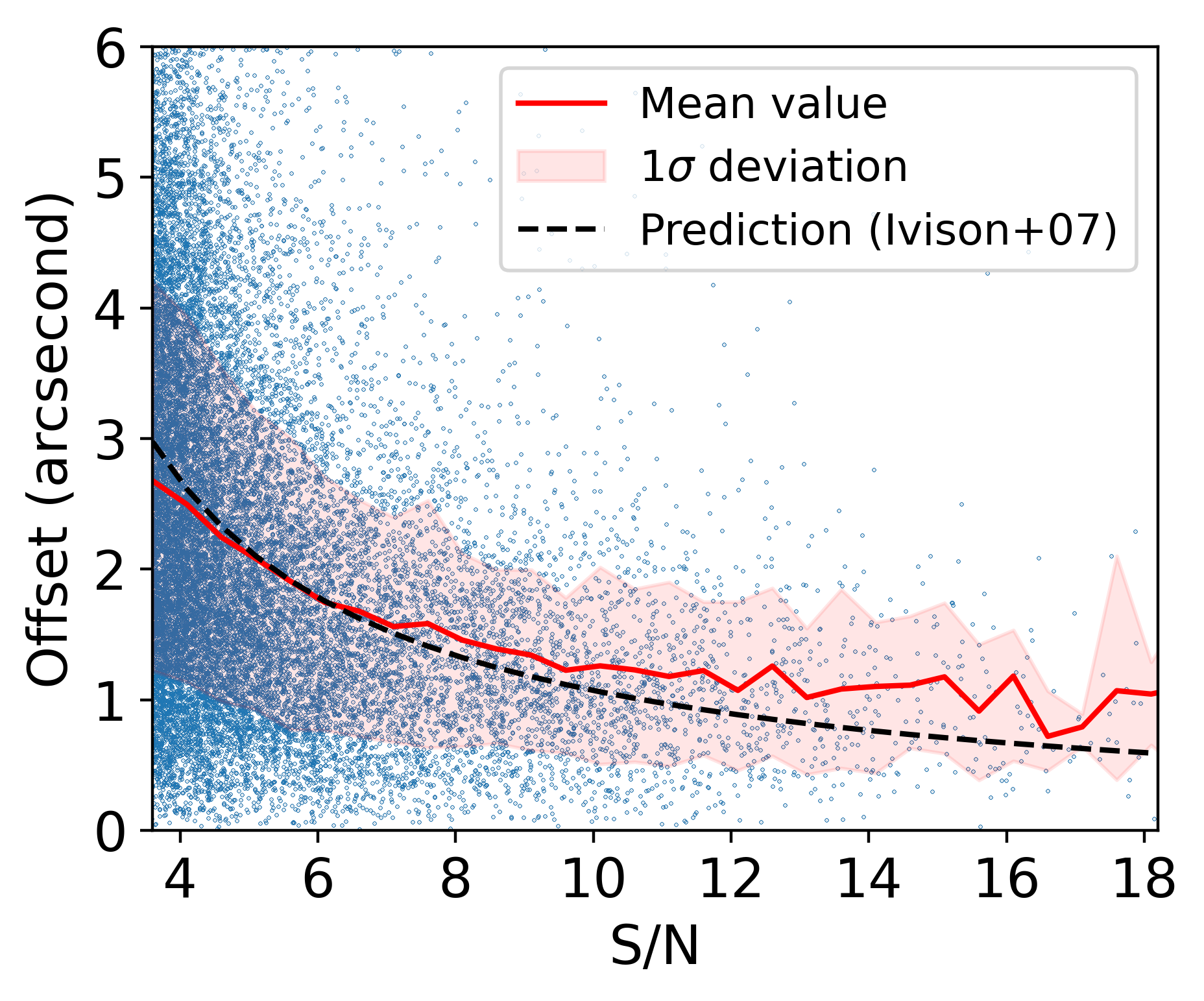}
    \caption{Positional offset between the positions of input sources and recovered sources as a function of the S/N in our simulations. The shaded region is the offset distribution from the 16th to 84th percentile, which can represent the 1$\sigma$ uncertainty. The offset drops as the S/N increases. For comparison, the prediction from \citet{Ivison2007} is also shown as the dashed line. }
    \label{fig:offset}
\end{figure}

The positional uncertainty primarily arises from instrumental and confusion noise, which is important to be considered when interpreting the precise positions of detected sources. We estimated this uncertainty by comparing the offsets between input sources and recovered sources. We calculated the offset as a function of the S/N using the same simulation results.
As illustrated in Fig.~\ref{fig:offset}, the positional offset tends to decrease as the S/N increases. Our estimated values
are consistent with the prediction from \citet{Ivison2007} but show a slight deviation for sources with a high S/N, which could be caused by pixelization.

\section{Ancillary data} \label{sec3}
We utilized ancillary data to aid our analysis, including $3\,{\rm mm}$ observations from MUSTANG-2 \citep{dicker2014}, $1.1\,{\rm mm}$ observations from AzTEC \citep{Wilson2008}, SPIRE, and Submillimeter Array (SMA) observations.

\subsection{MUSTANG-2 data}
We conducted $3\,{\rm mm}$ continuum observations of the 4C\,23.56 field using the MUSTANG-2 $90\,{\rm GHz}$ bolometer camera on the Green Bank Telescope (GBT) as part of program GBT21A-299 (PI: T.~Greve). These observations provide a resolution of 9\arcsec\ and an instantaneous field of view (FoV) of 4.2\arcmin\ (diameter). We employed a Daisy scanning pattern with a scanning radius of 3.5\arcmin. The 10 hours of telescope time were divided into two sessions in March 2021 and November 2021. Before each 20-minute scan, we conducted flux calibration observations using a nearby bright flux calibrator source. In total, the observations accumulated 4.7 hours of on-source time.

The raw data were recorded as time-ordered data from each responsive detector and subsequently calibrated and processed using the MUSTANG-2 MIDAS IDL pipeline \citep{Romero2020}. During processing, a Fourier filter was applied to reduce the RMS of the original time stream. Given the unstable atmospheric conditions, we employed a high-pass filter ($0.1\,{\rm Hz}$) to minimize atmospheric contamination. We applied the corresponding transfer function to account for signal loss during filtering. Finally, we smoothed the signal map with a Gaussian kernel (FWHM $\sim$\,9\arcsec). The RMS of the final product reaches ${\sim}\,21\,{\rm \mu Jy}$ at the center and ${\gtrsim}\,50\,{\rm \mu Jy}$ at the edge of the map. The footprint of the MUSTANG-2 map is outlined in pink in Fig.\,\ref{fig:ovlay}.

It is important to note that the $3\,{\rm mm}$ flux densities of our $850\,{\rm \mu m}$-selected sources are significantly lower than expected (by ${\sim}\,50\%$). It could be due to either steeper Rayleigh-Jeans slopes or the decrement from the Sunyaev-Zel'dovich (SZ) effect, which potentially introduces a larger uncertainty in our photometric redshift estimation. We emphasize the need for caution when constraining long-wavelength dust continuum with low-resolution data. Further details and interpretations of this issue are presented in Appendix~\ref{3mm}.

\subsection{ASTE AzTEC data}
The AzTEC $1.1\,{\rm mm}$ continuum map and source catalog of the 4C\,23.56 field were obtained as part of the AzTEC Cluster Environment Survey \citep[ACES;][]{Zeballos2018}, and kindly provided to us (Zeballos, private communications). These observations were conducted in a Lissajous pattern using the $10\,{\rm m}$ Atacama Submillimeter Telescope Experiment (ASTE) between September and November 2008, covering a ${\gtrsim}\,250\,{\rm arcmin}^2$ area with an angular resolution of ${\sim}\, 30\arcsec$ (FWHM). Its FoV is comparable to the cropped map of the SCUBA-2 data.
The total integration time is 35\,hours, and the central RMS reaches ${\sim}\,0.6\,{\rm mJy}$, which is ${\sim}\,4\times$ deeper than the confusion limit. The RMS increases to ${\gtrsim}\,1.2\,{\rm mJy}$ at the edge of the map. We refer to \citet{Zeballos2018} for additional details on the observations and the map properties.

\subsection{{\it Herschel} SPIRE data}\label{subsection:SPIRE}
We retrieved the SPIRE data from the Search for Protoclusters with {\it Herschel} (SPHer) survey \citep{Rigby2014} and the \textit{HErschel} Radio Galaxies Evolution (HeRGE) survey \citep{Drouart2014}, both of which are available in the ESA {\it Herschel} Science Archive.
We processed the archival data using the {\it Herschel} Interactive Processing Environment \citep[HIPE;][]{Ott2010}. The resulting RMS noise in the SPIRE maps is 2.5, 2.5, and $3\,{\rm mJy/beam}$ at 250, 350, and $500\,{\rm \mu m}$, respectively. The uncertainties in the SPIRE data are primarily driven by the confusion noise in the maps, which is 5.8, 6.3, and $6.8\,{\rm mJy/beam}$ at $250$, $350$, and $500\,{\rm \mu m}$, respectively \citep{Nguyen2010}. However, the SPIRE data are severely contaminated by the foreground galactic diffuse emission, which could potentially lead to inaccurate photometric measurements. The ``high-pass filter'' pipeline in HIPE has been applied to reduce such contamination.

To measure the SPIRE photometry, we utilized the time-line fitting algorithm on the time-ordered data with prior information from the $850\,{\rm \mu m}$ positions, which is considered the most reliable method for the SPIRE photometry measurement \citep{Pearson2014}. For the $850\,{\rm \mu m}$ sources not robustly detected in SPIRE (i.e., ${<}\,2\sigma$), we attempted to constrain their flux density by simultaneous fitting all SCUBA-2 sources to the SPIRE images using SUSSEXtractor with a threshold of 2$\sigma$ on the source positions as the prior \citep{Savage2007}. In cases of a non-detection, we recorded the pixel value of the centroid positions of $850\,{\rm \mu m}$ sources and the corresponding standard deviation of the nearby $5\times5$ pixels for further analysis. It is worth noting that the image-based extraction method can lead to an underestimation of point source signals due to pixelization of the time-ordered data. To address this effect, we corrected our 250, 350, and $500\,{\rm \mu m}$ photometric measurements by 4.9, 6.9, and 9.8\%, respectively \citep{Rigby2014}.

\subsection{SMA data}
High-resolution interferometric observations at $345\,{\rm GHz}$ and $400\,{\rm GHz}$ were conducted with SMA under the program SMA2021A-A011 (PI: C.C.\,Chen, 2021DDT). The seven brightest SMGs in the field of 4C\,23.56 were targeted in the subcompact configuration, resulting in a resolution of ${\sim}\,3\arcsec$.
The observations were carried out in September 2021, under atmospheric conditions corresponding to a precipitable water vapor of $2.5{\rm mm}$. We used two tracks (12 hours) with seven antennas and with ${\sim}\,2$\,hours of on-source time in total. Dual-receiver mode was employed with 345Rx and 400Rx receivers centered at $334\,{\rm GHz}$ and $405\,{\rm GHz}$, respectively. Calibration was performed using CASA (v5.7.0) with 2025+337 as the phase calibrator, MWC349A as the flux calibrator, and either BLLAC or CALLISTO as the bandpass calibrator. Images were created using natural weighting, and the CLEAN algorithm was applied in regions with signal-to-noise ratios greater than 3. The resulting average RMS for each pointing is ${\sim}\,2.0-2.5\,{\rm mJy/beam}$.

\section{Analysis} \label{s4}

In this section we evaluate the number counts and the SCUBA-2 source overdensity to determine if there is a excess of SMGs in the 4C\,23.56 field.

\subsection{Number counts} \label{number count}

To estimate the differential number counts, we first corrected the contribution from each source with an observed flux $S_{\rm obs}$ by the effective area and fidelity:
\begin{equation}
n_i = \frac{f_{\rm fid}\,(S_{{\rm obs}, i})}{A_{\rm eff}\,(S_{{\rm obs}, i})},
\end{equation}
where $n_i$ is the effective number contribution for a given source $i$, $f_{\rm fid}$ is the fidelity, $A_{\rm eff}$ is the effective area, which is defined as the area of the map within which a source with flux $S_{\rm obs}$ can be detect above 3.5$\sigma$. We then de-boosted the flux of each source following the Gaussian probability distributions defined in Sect. \ref{sec:deboosting} to get their intrinsic fluxes $S_{\rm in}$.
These are used to calculate the contributions to the number counts from source $i$, which is expressed as
\begin{equation}
    \xi_i\,(S_{{\rm in}, i}) = \frac{n_i}{C\,(S_{{\rm in}, i})},
\end{equation}
where C is the completeness.
We summed up the contribution from each source to calculate the differential number counts as
\begin{equation}
    \frac{dN}{dS}\,(S_{\rm in}) = \frac{\Sigma\,\xi_i\,(S_{\rm in})}{\Delta\,S_{\rm in}},
\end{equation}
where
$\Sigma\,\xi_i\,(S_{\rm in})$ is the sum of the contributions from all sources within the flux bin $S_{\rm in} \pm \Delta\,S_{\rm in}/2$ (we adopt $\Delta S_{\rm in}$\,=\,3\,mJy in this work).
To account for the uncertainties from de-boosting, we repeated this process 5,000 times and measured the mean values and the standard deviation of the number counts per flux bin, which gives the estimated number counts of the studied region.

\begin{figure}[t]
    \centering
    \includegraphics[width=0.42\textwidth]{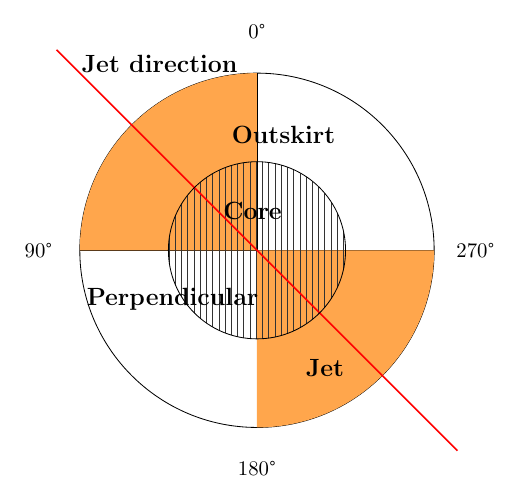}
    \caption{Illustration of the H$z$RG environment. We separate the region by the distance to the H$z$RG and the angle to the radio jet. The regions close to the jet are defined as the jet subsample and the rest as the perpendicular subsample. The position angles are indicated on the diagram. }
    \label{fig:schematic}
\end{figure}

It has been suggested that SMGs are usually concentrated within the central $\sim$\,2\,Mpc of H$z$RGs \citep[e.g.,][]{Ivison2000,Greve2007,Hatch2011,Zeballos2018}. In addition to the region within the entire map, we also estimated the number counts of the central region $\sim$\,2\,Mpc (4\arcmin\ radius aperture at $z\,{\sim}\,2.5$) and outskirts (annulus with $4\arcmin\,{\leq}\, r\,{<}\,9\arcmin$) regions of the H$z$RG. For a visual clarity, a schematic diagram is shown in Fig.~\ref{fig:schematic}.

\begin{figure}[!t]
    \centering
    \includegraphics[width=0.48\textwidth]{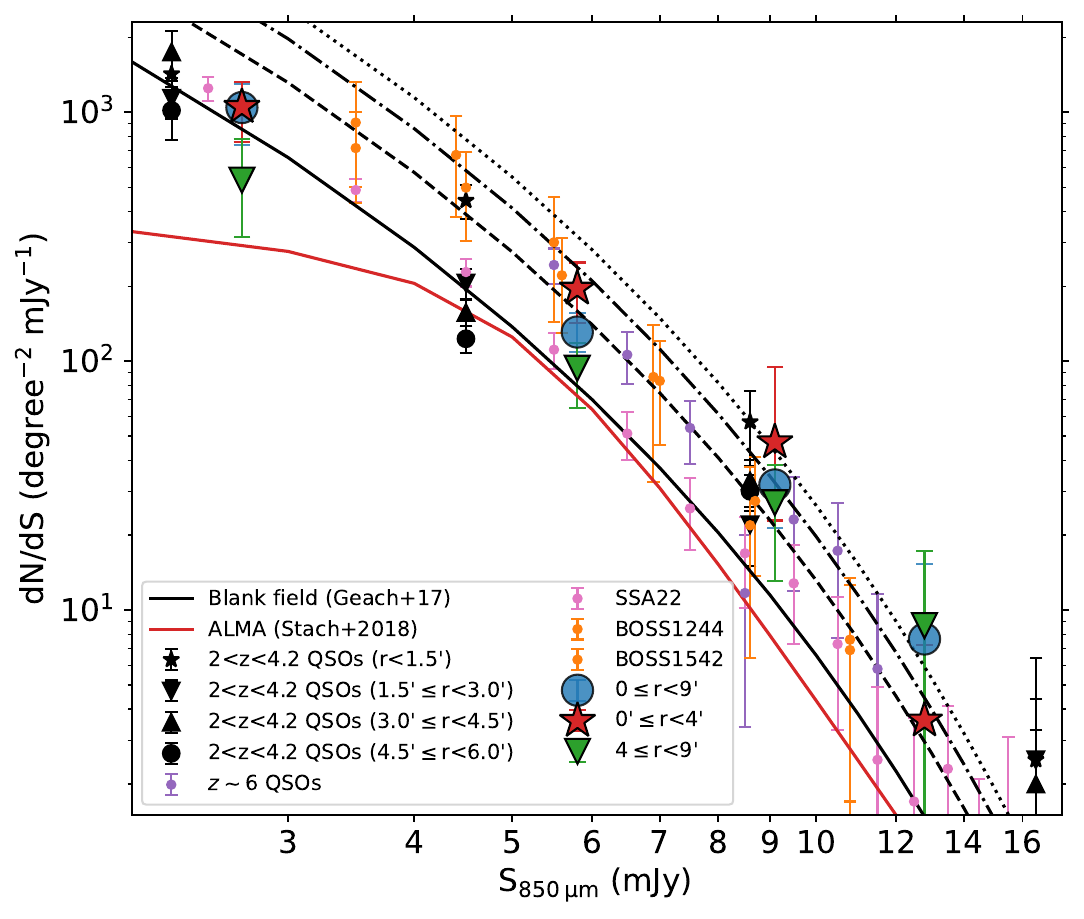}
    \caption{Differential number counts obtained within core and outskirts regions, with 3\,mJy as the bin size of $S_{850\,{\rm \mu m}}$.
    We show the blank field number counts from \citet[solid black curve]{Geach2017} and from \citep[solid red curve]{Stach2018} as reference. The blank field number counts multiplied by factors of 2, 3, and 4 are shown as the dashed, dash-dotted, and dotted curves, respectively. The number counts from the literature are shown for comparison \citep{Zhang2022,Li2023,AB2023,Zeng2024}. }
    \label{fig:nc}
\end{figure}

The number counts of these different regions and of the blank field from \citet{Geach2017} are shown in Table\,\ref{tab:number count} and Fig.\,\ref{fig:nc}. For comparison, we also present the blank field number counts multiplied by factors of 2, 3, and 4.

\begin{table}[tb]
    \caption{Differential number counts computed in different regions and overall cumulative number counts with corresponding overdensities.}
    \label{tab:number count}
    \centering
    \renewcommand{\arraystretch}{1.5}
    \scalebox{0.9}{
    \begin{tabular}{ccc|ccc}
    \hline
       Region & $S_{850\,{\rm \mu m}}$  &  d$N$/d$S$ & $S'_{850\,{\rm \mu m}}$ & $N$(>\,$S'$) & $\delta$ \\
       &{\small(mJy)} & {\small(deg$^{-2}$mJy$^{-1}$)} & {\small(mJy)}& {\small(deg$^{-2}$)} \\
       \hline
        \multirow{4}{*}{$r\,{<}\,4\arcmin$}&2--5&1052.0$^{+289.1}_{-267.8}$ & 2&2716.0$_{-549.0}^{+602.6}$&0.1$^{+0.2}_{-0.2}$\\
        &5--8& 195.8$^{+54.0}_{-53.4}$ &3&1123.5$_{-161.0}^{+176.5}$&0.1$^{+0.2}_{-0.1}$\\
        &8--11 & 47.3$^{+24.4}_{-47.3}$&4&625.6$_{-63.3}^{+74.5}$&0.2$^{+0.2}_{-0.1}$\\
        &11--14 &3.6$^{+3.6}_{-3.6}$ &5&425.3$_{-51.1}^{+46.2}$&0.6$^{+0.1}_{-0.2}$\\
        \cline{1-3}
        \multirow{4}{*}{$4\arcmin\,{\leq}\,r\,{<}\,9\arcmin$}& 2--5 &533.8$^{+218.9}_{-249.1}$&6&273.5$_{-43.0}^{+43.6}$ &0.8$^{+0.3}_{-0.3}$\\
        & 5--8 &93.7$^{+28.6}_{-24.9}$ &7&163.5$_{-35.2}^{+31.9}$&0.9$^{+0.4}_{-0.4}$\\
        & 8--11&27.0$^{+13.9}_{-11.4}$ &8&95.8$_{-27.1}^{+26.3}$&1.0$^{+0.6}_{-0.5}$\\
        & 11--14&8.7$^{+9.0}_{-8.7}$ &9&55.2$_{-25.5}^{+17.2}$&1.1$^{+0.6}_{-1.0}$\\
        \cline{1-3}
        \multirow{4}{*}{$r\,{<}\,9\arcmin$}& 2--5 &1045.2$^{+306.1}_{-260.7}$ &10 &32.1$_{-18.0}^{+10.3}$&1.2$^{+0.7}_{-1.2}$\\
        & 5--8&130.9$^{+21.9}_{-24.9}$ & 11 & 16.5$_{-16.5}^{+11.8}$ & 0.9$^{+1.4}_{-1.9}$\\
        & 8--11 & 31.8$^{+10.5}_{-11.8}$ & 12 & 4.3$_{-4.3}^{+9.9}$ & -0.2$^{+1.8}_{-0.8}$\\
        & 11--14 & 7.6$^{+6.5}_{-7.6}$ & 13 & 1.2$^{+1.2}_{-1.2}$ & -0.6$^{+0.3}_{-0.4}$\\
        \hline
    \end{tabular}
    }
    \tablefoot{Uncertainties are the standard deviation of the number counts in each flux bin after 5000 realizations with de-boosting process following the Gaussian probability distributions described in Sect.\,\ref{number count}. Due to the limited area, we calculated the average differential number counts across 3\,mJy.}
\end{table}

\subsection{Overdensity analysis} \label{ovd}

To further quantify the apparent excess of SMGs observed in the vicinity of 4C\,23.56 compared to the field, we made use of the well-known projected overdensity estimator:
\begin{equation}
\delta=\frac{N - \langle N \rangle}{\langle N \rangle},
\label{eq:ovd}
\end{equation}
where $N$ is the observed source counts and $\langle N \rangle$ is the expected blank field counts.

Several works have assessed whether the SMGs in the vicinity of the H$z$RG have a preferential alignment and are spatially correlated with the radio jet \citep[e.g.,][]{Stevens2003,Zeballos2018}. To examine the spatial distribution of SMGs in detail, we
generate da set of mock maps using the blank field number counts from \citet{Geach2017} with a matching RMS noise level as our reference fields, and used Eq.~\ref{eq:ovd} to compute overdensities in different locations, instead of deriving number counts.
{Since it directly compares detected sources without any corrections, it does not depend on the de-boosting, completeness, or fidelity process.
Compared to the overall number counts, this method offers a better measure of spatial overdensities. }

In the previous section we estimated the SMG overdensity within an inner ($r\,{<}\,4\arcmin$) and outer ($4\arcmin\,{\leq}\,r\,{<}\,9\arcmin$) region. In this section we further explore its variation as a function of flux and location in the HzRG environment.

\begin{figure}[t]
    \centering
    \includegraphics[width=0.47\textwidth]{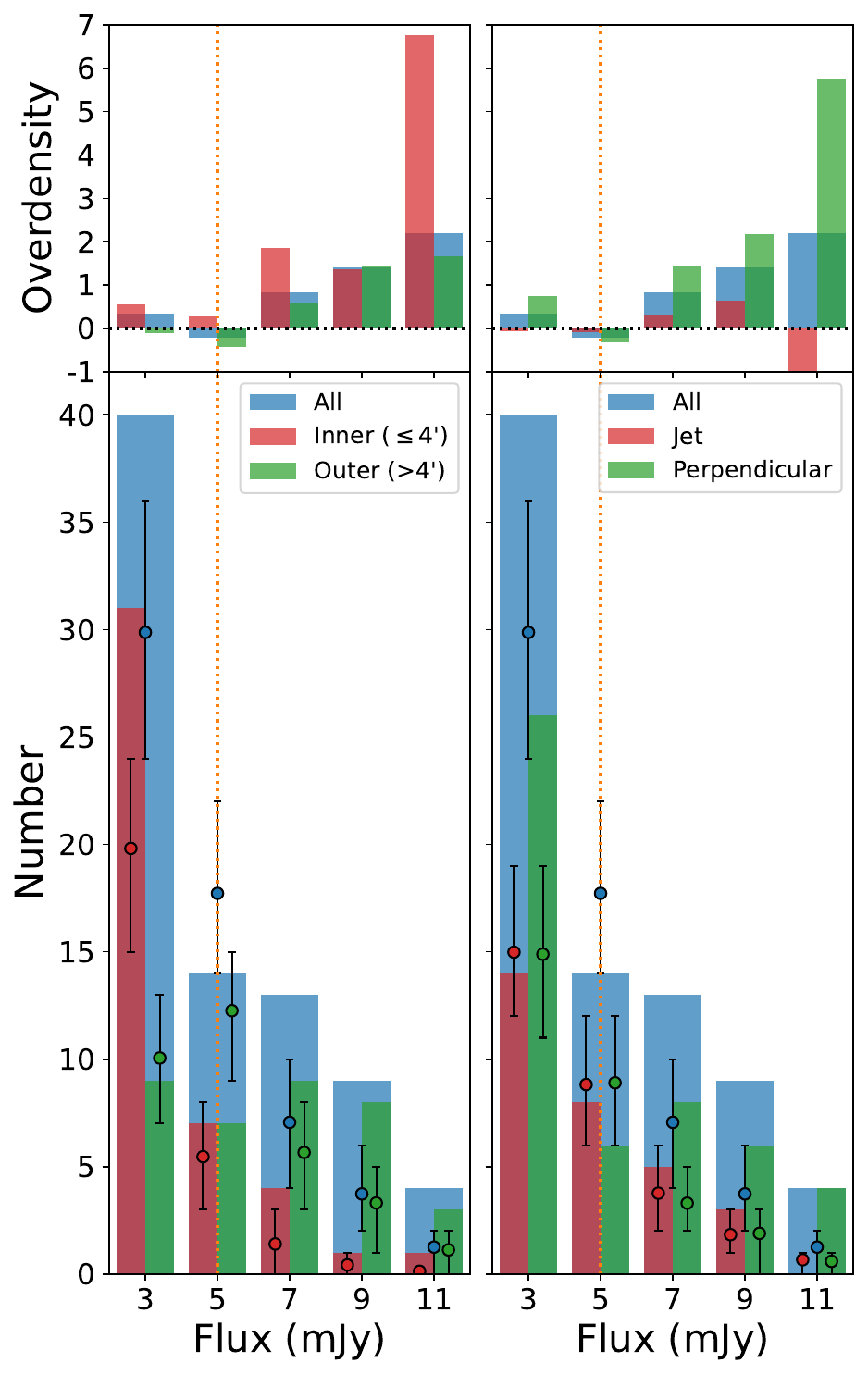}
    \caption{Number distribution of SMGs as a function of the flux density.
    Points with error bars represent the expected numbers of SMGs from the blank field, while the histograms are the measured numbers in the 4C\,23.56 field.
    The blue histogram illustrates the total number of SMGs within flux bins of 2\,mJy. \textit{Left:} SMG sample divided into the central 4\arcmin\ from the H$z$RG (inner; red) and the outer ${>}\,4$\arcmin (outer; green) regions.
    \textit{Right:} SMG sample based on angular distribution: within $45^{\circ}$ of the jet direction (red) outside the jet (perpendicular; green). The corresponding overdensities as functions of flux are plotted in the top panel on each plot. The dotted orange line indicates the flux threshold (i.e., 5\,mJy) that we used for further analysis. Overdensity increases with the flux density in both inner and outer regions. But neither the overdensity nor this positive correlation exists in the jet region.
    }
    \label{fig:ovdf}
\end{figure}

\begin{figure}[t]
    \centering
    \includegraphics[width=0.47\textwidth]{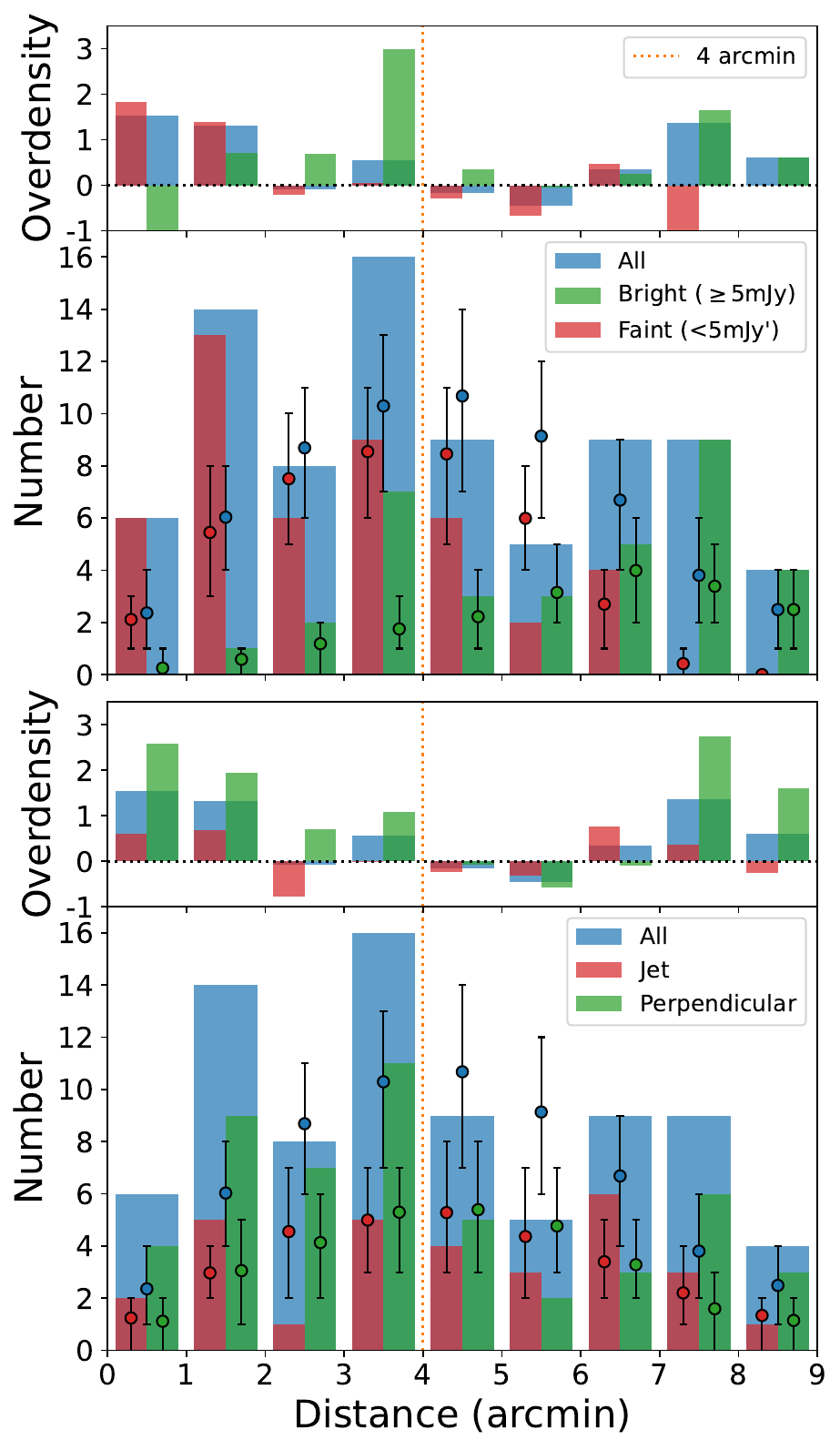}
    \caption{Number distribution of SMGs as a function of the radial distance to the center.
    The points with error bars are the expected numbers of SMGs from the blank field and the bars are the measured numbers in the field. The blue histogram shows the total number of SMGs within each bin.
    \textit{Top:} SMG sample divided into bright ($S_{850\,{\rm \mu m}}\,{\geq}\,5$\,mJy; green) and faint ($S_{850\,{\rm \mu m}}\,{<}\,5$\,mJy; red) subsamples.
    \textit{Bottom:} SMG sample divided into jet (green) and perpendicular (red) region SMGs (see the schematics in Fig.~\ref{fig:schematic}).
    The corresponding overdensities are displayed as lines in the same colors. The faint SMGs show an excess within the central 2\arcmin, while overdensity of bright SMGs peaks at $3\arcmin\,{\leq}\,r\,{<}\,4\arcmin$. Overdensity in jet region shows less dependence on the distance compared to the perpendicular region.
    }
    \label{fig:ovdr}
\end{figure}

\begin{figure}[t]
    \centering
    \includegraphics[width=0.48\textwidth]{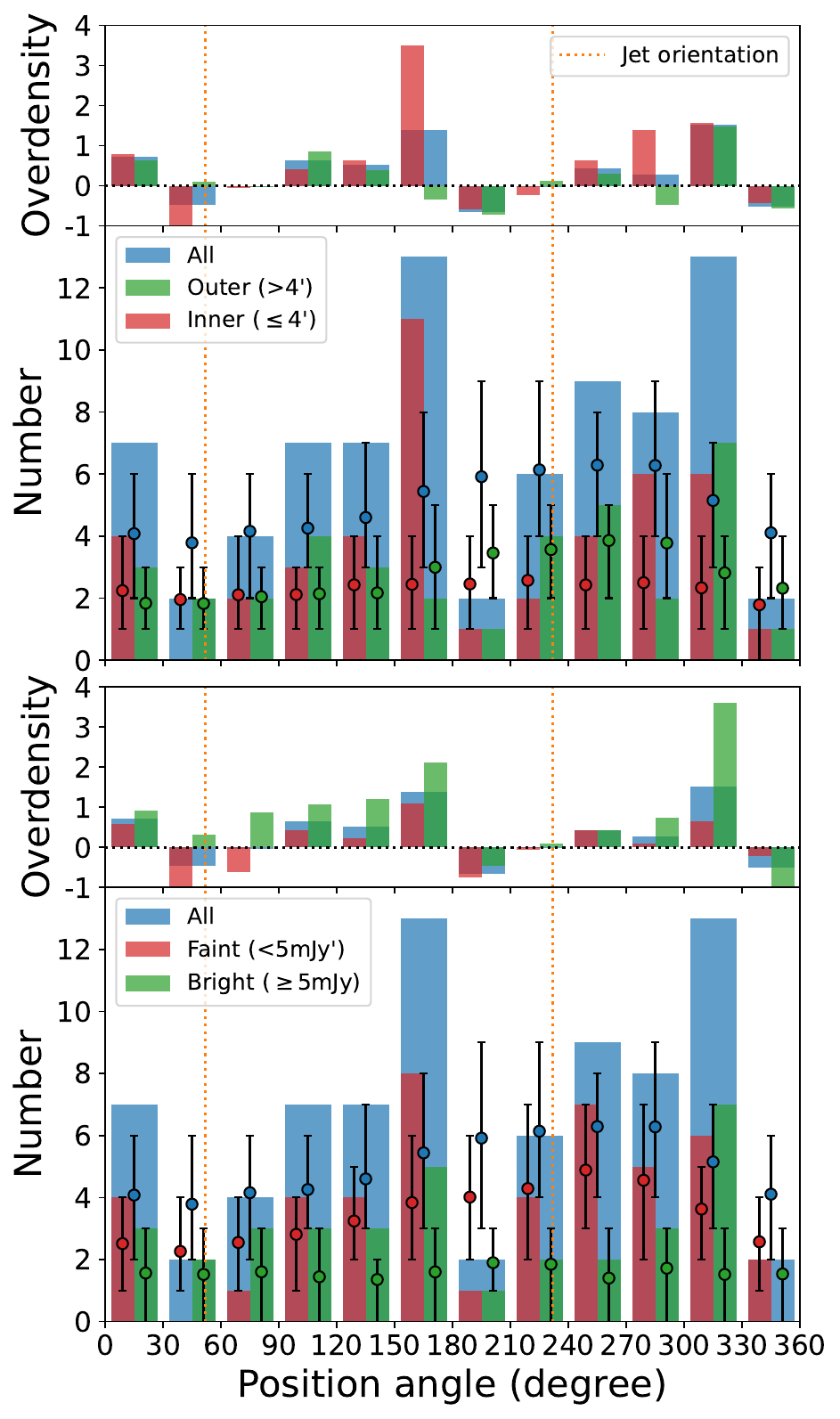}
    \caption{Number distribution of SMGs as a function of the position angle.
    The fluctuation of the simulated SMG number is due to the asymmetric sensitivity of the SCUBA-2 data.
    The top panels in each plot show the corresponding overdensities as lines in the same colors. And the dashed orange lines indicate the rough direction of the radio jet. The angular dependence is depicted for both bright and faint SMGs in both inner and outer regions.
    }
    \label{fig:ovda}
\end{figure}

Figures\,\ref{fig:ovdf},\,\ref{fig:ovdr}, and \ref{fig:ovda} show the variation in the number counts of SMGs, and their corresponding overdensity, as a function of flux density, radial angular distance, and positional angle (PA) to the H$z$RG, respectively. The corresponding blank field number counts and their standard deviations are shown as open circles  with error bars. In each figure we further classify the SMGs into ``inner'' versus ``outer,'' ``jet'' versus ``perpendicular'' (``per''), and ``bright'' versus ``faint'' subsamples. The jet sample consists of SMGs within 45 degrees of the radio jet direction, while the perpendicular sample constitutes the SMGs that lie in the perpendicular direction within 45 degrees (see Fig.\,\ref{fig:schematic}).

In Fig.~\ref{fig:ovdf} we see that the SMG overdensity is positively correlated with $850\,{\rm \mu m}$ flux. This correlation is more obvious in the inner region than in the outer region, while it does not exist in the jet region. Based on their overdensities, we used 5\,mJy as the threshold to separate the SMGs into faint and bright samples. In Fig.~\ref{fig:ovdr}, the faint samples demonstrate a preference for the inner $2\arcmin$ region, whereas the overdensity for bright SMGs peaks at $3\arcmin\,{<}\,r\,{<}\,4\arcmin$ and becomes lower in the inner $2\arcmin$ region, which cannot be found in the jet region. Figure~\ref{fig:ovda} shows that SMGs exhibit a strong angular preference, which is also more prominent among bright SMGs. Overdensity reaches its maximum in the direction perpendicular to the radio jet, with fewer sources along the jet direction.

Our overdensity analysis suggests the presence of more bright SMGs in the protocluster field than the blank field, with SMGs predominantly distributed perpendicular to the jet direction.
This angular preference exists for both faint and bright SMGs (Fig. \ref{fig:ovda}), indicating a potential association of both populations with the system.

\subsection{Photo-\texorpdfstring{$z$}{z} support for an anisotropic distribution of SMGs in \texorpdfstring{4C\,23.56}{4c23}}

To test whether the anisotropy is caused by an obvious projection effect and if bright SMGs are more likely to be associated with the structure, we utilized multiwavelength data to derive the photo-$z$ for all $850\,{\rm \mu m}$-selected sources. Due to the poor angular resolution of the SCUBA-2 observation and relatively shallow ancillary data in the optical and near-infrared regimes, cross-matching $850\,{\rm \mu m}$ sources with optical counterparts is challenging. In this work, we only attempted to derive photo-$z$ from the FIR photometry.

\begin{figure}[t]
    \centering
    \includegraphics[width=0.47\textwidth]{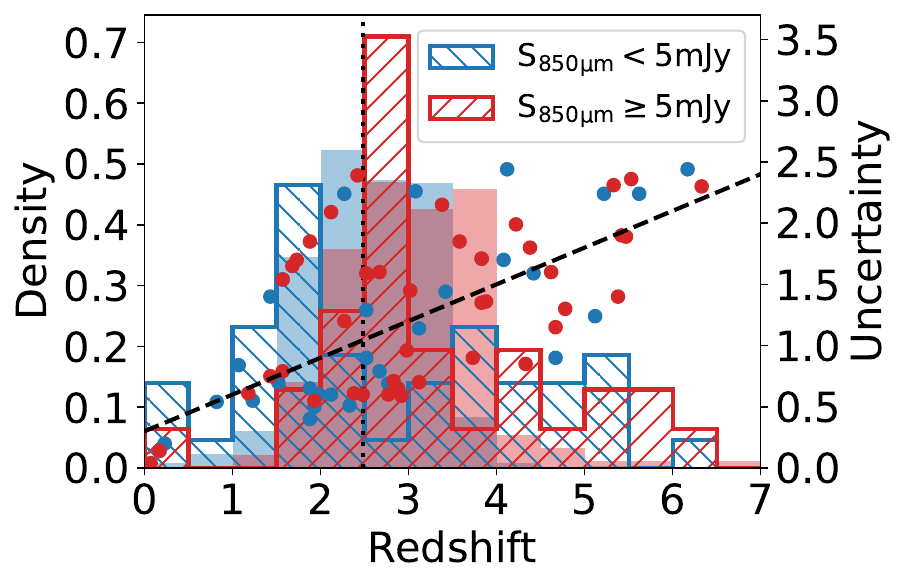}
    \caption{Photo-$z$ density distribution of the 80 SMGs detected above $3.5\sigma$ in the 850$\,{\rm \mu m}$ map (hatched) and the corresponding SMG redshift distribution of the blank field from \citet{Dudzev2019} as the background. The dotted line represents the redshift of 4C\,23.56. We divide the sample into bright (${\rm S}_{\rm obs}$\,$\geq$\,$5{\rm mJy}$) and faint (${\rm S}_{\rm obs}$\,$<$\,$5{\rm mJy}$) SMGs. The photo-$z$ uncertainties are indicated as scatters with corresponding colors and its relation with the redshift ($\Delta z\,{\approx}\,0.3(1+z)$) is shown by the dashed line. Compared to the redshift distribution of field SMGs, bright SMGs show a higher density at the redshift of the H$z$RG, while faint sources are concentrated on a lower redshift.}
    \label{fig:redshift}
\end{figure}

Table~\ref{multi-z} lists the photometric redshifts determined by the MMP$z$ code \citep{casey2020}. MMP$z$ can provide relatively satisfactory redshift solutions with FIR photometry ($\Delta z/(1+z)$\,$\approx0.3$) based on prior knowledge about the peak of the dust spectral energy distribution (SED). The peak wavelength are thought to be correlated with the infrared luminosity, as dust is thought to be hotter in galaxies with higher star formation rates \citep[the so-called $\lambda_{\rm peak}$ technique;][]{casey2018}. MMP$z$ has shown the capability of providing rough redshift estimates in absence of expensive spectroscopic observations \citep[e.g.,][]{Montana2021,Cooper2022}.
The redshift distribution of SMGs in the 4C\,23.56 field is shown in Fig.~\ref{fig:redshift}. The redshift distribution of the field SMGs is also shown for reference \citep{Dudzev2019}. The estimated photometric redshifts have relatively large uncertainties, which are not enough to confirm protocluster memberships and can be only used to exclude SMGs that are less likely associated with the structure.

\begin{figure}[t]
    \centering
    \includegraphics[width=0.49\textwidth]{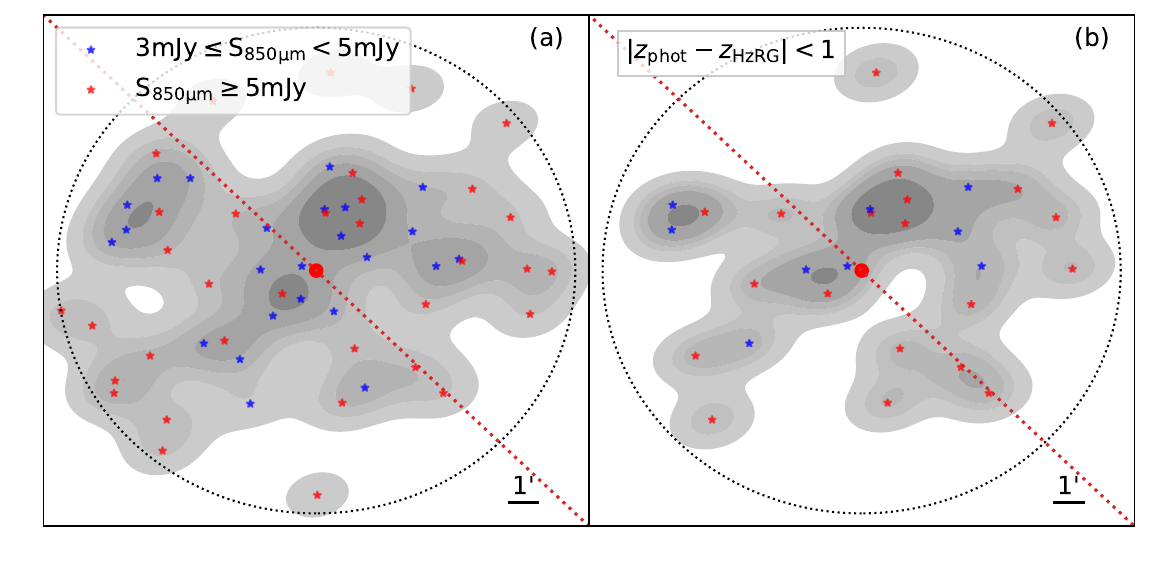}
    \caption{Kernel density estimate plots of (a) SCUBA-2 $850\,{\rm \mu m}$ and (b) $850\,{\rm \mu m}$ sources with photo-$z$ close to 4C\,23.56 ($\Delta z\,{<}\,1$),
    with 10\% as the contour step. The direction of the radio jet is shown as dotted red lines. The FoV
    is indicated as a dotted circle.
    Bright and faint SCUBA-2 sources are color-coded by red and blue, respectively. After the redshift selection, the SMGs still show the angular preference.
    }
    \label{fig:fird}
\end{figure}

The photo-$z$ determined by MMP$z$, shows that 29 candidates have photo-$z$ similar to that of the H$z$RG (with $\Delta z\,{<}\,1$). As indicated in Fig.~\ref{fig:fird}, the anisotropic distribution still remains when only considering these 29 sources. Thus, the scenario that protocluster SMGs in the 4C\,23.56 field have an angular preference is further strengthened.

In addition, to see if bright SMGs are more likely associated with the H$z$RG system, we plot the photo-$z$ distributions of the bright and faint samples separately.
The bright SMGs have photo-$z$ closer to $z\,{=}\,2.5$ and the photo-$z$ of faint SMGs shows a concentration at a lower redshift, which also supports the idea that bright SMGs are more likely to be associated with the structure.

One caveat of the result is that this technique could potentially cause a bias toward higher redshift for luminous sources. In this case, the difference in the redshift distributions of the bright and faint SMGs can be explained by the artifact introduced by the "$\lambda_{\rm peak}$" technique.

\section{Discussion}\label{s5}
\subsection{Spatial distribution of SMGs in AGN environments}

To examine if the spatial distribution of SMGs around 4C\,23.56 is common for AGN environments, we made a comparison with other AGN fields.
Surveys of AGNs with similar properties, such as enormous Ly$\alpha$ nebulae (ELANe) and other H$z$RG fields, find a similar angular preference \citep{Zeballos2018,AB2023}. To check if the spatial distributions of SMGs in those studies are consistent with the anisotropic distribution found in the 4C\,23.56 field. We selected eight H$z$RG and ELAN fields at $2\,{<}\,z\,{<}\,3$ from the literature for further analysis \citep[see our Table~\ref{tab:pc} as well as][]{Zeballos2018,Nowotka2022,AB2023}.

\begin{figure}[t]
    \centering
    \includegraphics[width=0.47\textwidth]{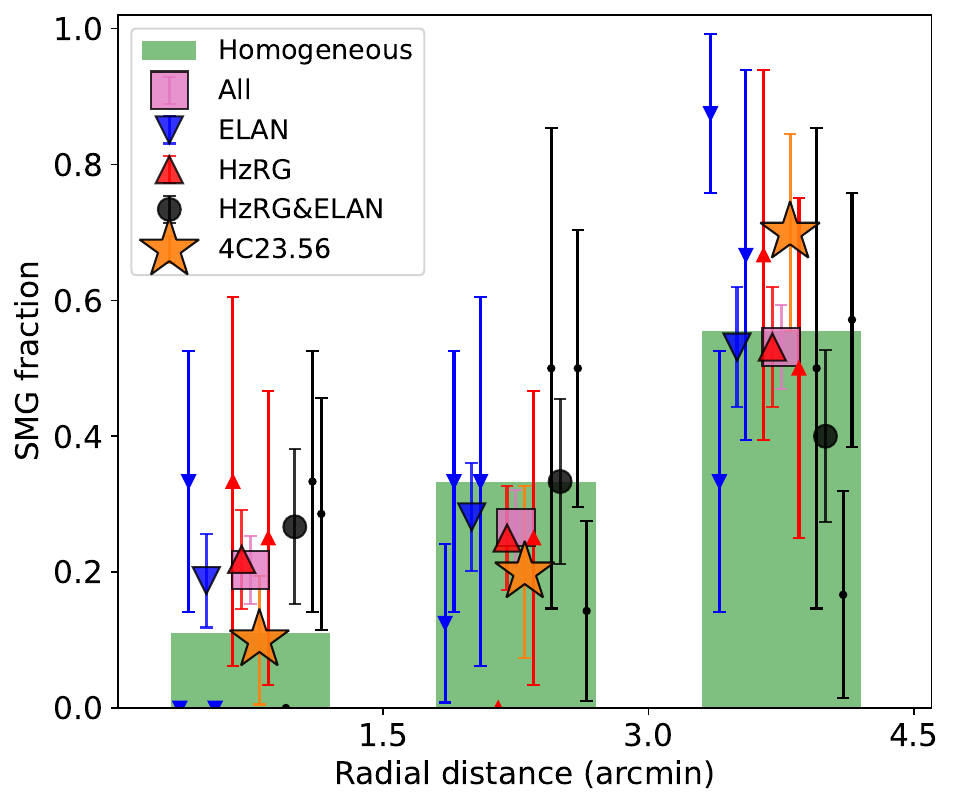}
    \caption{Comparison of the radial variation in the SMG fraction. The green bar indicates the expected SMG fraction for a homogeneous distribution.
    The small symbols are the values for each individual field. The large symbols are the stacked results for corresponding types of fields. Poisson noise is assumed as the associated uncertainty. We can see that SMGs are concentrated in the central 1.5\arcmin\ in general, while more SMGs can be from at 3\arcmin\,${\leq}\,r\,{<}\,$4.5\arcmin\ in the 4C\,23.56 field.}
    \label{fig:compare_r}
\end{figure}

In the 4C\,23.56 field, bright SMGs tend to avoid the core region, and faint SMGs are concentrated in the central 2\arcmin (see Sect. \ref{s4}). This is in contrast to previous reports that SMGs only show a clear excess within $r\,{<}\,1.5$\arcmin \citep{Zeballos2018,AB2023}. Considering the different sensitivity of these surveys, we only included bright sources with high completeness fraction ($S_{850\,{\rm \mu m}}\,{\geq}\,5$\,mJy or $S_{\rm 1.1mm}\,{\geq}\,3.4$\,mJy) for our comparison.
We used the catalogs from literature and calculated the number of bright SMGs within the radial distance $r\,{\leq}\,1.5\arcmin$, $1.5\arcmin\,{<}\,r\,{\leq}\,3\arcmin$, and $3\arcmin\,{<}\,r\,{\leq}\,4.5\arcmin$ \citep{Zeballos2018,Nowotka2022,AB2023}. For a fair comparison, we normalized the source counts by the total number of bright SMGs within the central 4.5\arcmin\ in each field (SMG fraction, Fig.~\ref{fig:compare_r}). In general, we find that the bright SMGs are concentrated in the central 1.5\arcmin\ region. For central AGNs classified as both H$z$RGs and ELANe (H$z$RG\&ELAN), the SMG fraction is higher in the central 1.5\arcmin\ and decreases at $3\arcmin\,{<}\,r\,{\leq}\,4.5\arcmin$. Interestingly, the 4C\,23.56 field does not show an excess of bright SMGs in the central 3\arcmin, but rather at $3\arcmin\,{<}\,r\,{\leq}\,4.5\arcmin$. This might suggest that the field of 4C\,23.56 is under a different evolutionary stage, than the control samples.

\begin{figure}[t]
    \centering
    \includegraphics[width=0.47\textwidth]{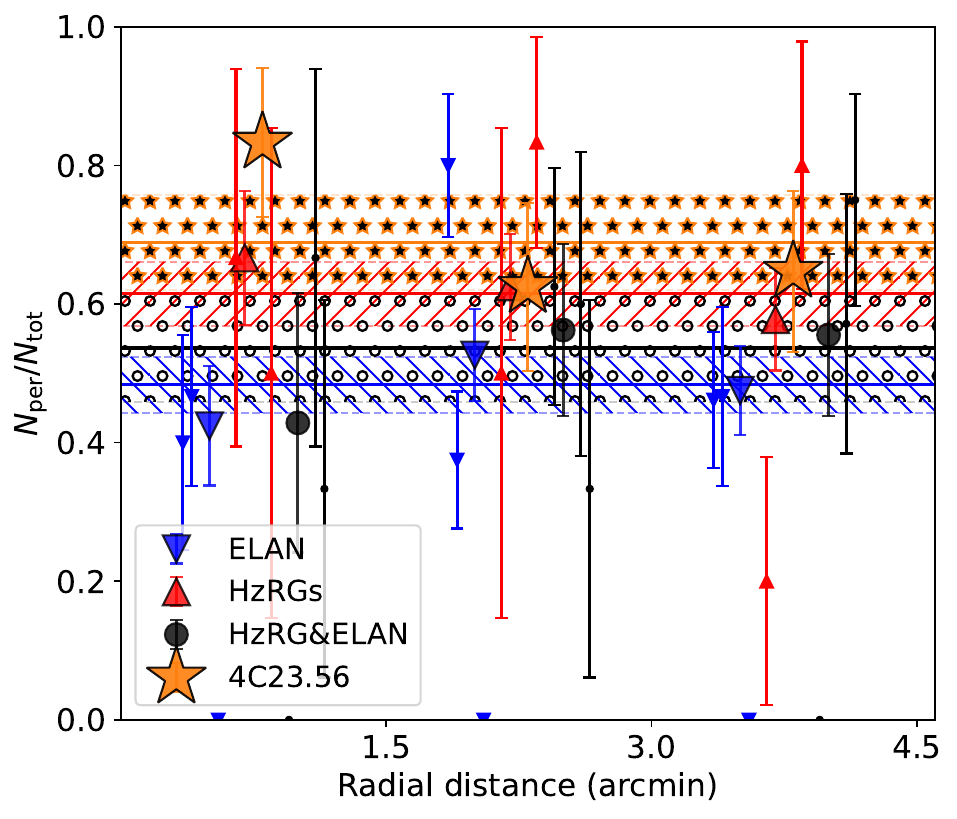}
    \caption{Comparison of the radial variation in the anisotropic level of the SMG distribution. All SMGs are included in this analysis. The small symbols are the values for each individual field. The large symbols are the stacked results for corresponding categories. The horizontal lines and hatches in different colors indicate the average values and the associated Poisson uncertainties of each AGN field. We can see that in the 4C\,23.56 field, although SMG fraction is consistent with a homogeneous distribution at $r\,{<}\,1.5\arcmin$, SMGs show a stronger preference for the perpendicular region than other fields.}
    \label{fig:compare_a}
\end{figure}

To further quantitatively assess the angular preference of SMGs in the 4C\,23.56 and other AGN fields in a similar manner, we studied their fractions of SMGs in the perpendicular region within different annulus. For the ELAN fields, the perpendicular region represents the PA from the major axis of ELAN larger than 45$^\circ$. As shown in Fig.~\ref{fig:compare_a}, the average value for the ELAN fields is close to unity. According to this assessment, there is no clear evidence that SMGs in any ELAN fields have a dependence on the major axis of the Ly$\alpha$ nebula. For the H$z$RG\&ELAN fields, there is a small SMG excess along the jet direction at $r\,{<}\,1.5\arcmin$. The aspherical level of H$z$RG fields is more significant in the inner region and becomes less obvious as distance increases, which has the same trend as in the 4C\,23.56 field. If it is not caused by the large uncertainty introduced by the small number statistic errors, then this aspherical distribution can either be due to contamination from the increasing number of field SMGs in a larger area or the fact that the AGN feedback has less of an impact on more distant regions.

Such a jet-dependent anisotropic distribution has also be found in UV-selected galaxies in H$z$RG environments \citep{west1994,Kurk2004,Venemans2007,Uchiyama2022}. This angular preference has been explained by the correlation between the geometry of the cosmological structure and galaxy spin \citep{Peebles1969}. According to the tidal torque theory, massive galaxies favor spins perpendicular to the filament axis \citep{Codis2018}. The correlation can be reproduced if the radio jet is consistent with the spin direction and SMGs distribute along the dark matter filament. However, misalignment is commonly observed between radio jets and galaxy spins \citep[e.g.,][]{Wu2022}, which can arise naturally in high-resolution simulations \citep{Fanidakis2011,Hopkins2012}. Furthermore, the large-scale baryon distribution can be highly influenced by the AGN feedback and SMGs prove even less revealing of the dark matter overdensities due to their rarity \citep[e.g.,][]{Chapman2009,vDaalen2011,Miller2015}. In the 4C\,23.56 field, the same angular preference cannot be found in other galaxy populations \citep{Knopp1997,Mayo2012,Galametz2012,Lee2017}, which disfavors this scenario.

Alternatively, AGN feedback is equally able to cause the angular conformity \citep[e.g.,][]{Kauffmann2015,MartinNavarro2021}. A radio AGN is likely to efficiently heat surrounding gas anisotropically and shape the structure of the hot intracluster medium \citep{Barnes2019,Thomas2022,Bennett2023, Dong2023,Chapman2024}, which subsequently prevent galaxies along the radio jet from accreting the surrounding cold gas, or determines the distribution of gas filaments and reflected by the spatial distribution of SMGs \citep[e.g.,][]{Russell2019,Alberts2022,emonts2023,Emonts2023b}. But it is unclear if the jet can have a significant influence on megaparsec scales. Due to limited sample size in this study and the complexity of the AGN feedback and radio jet, we cannot draw a firm conclusion about SMG distribution around H$z$RGs in this pilot study. A systematic detailed study is reserved to strengthen this idea in the following RAGERS survey.

\begin{table*}[t]
    \caption{Overview of ELAN and H$z$RG fields at $2\,{<}\,z\,{<}\,3$.}
    \label{tab:pc}
    \centering
    \begin{tabular}{llcccccll}
    \hline
    \hline
        Name & Type & $z$ & Size & PA & N$_{\rm SMG}$ & RMS & Instrument & Reference\\
         & & & [kpc] & [deg] & & [mJy] & & \\
         \hline
        4C\,23.56 & H$z$RG(LAB) & 2.48 & 86 & 52 & 17 & 0.6 & SCUBA-2 & this work\\
        Jackpot & ELAN & 2.04 & 300 & -77 & 11 & 0.5 & SCUBA-2 &  \citet{AB2023}  \\
        MRC 2048-272 & H$z$RG(LAB) & 2.06 & 71 & 45 & 4 & 0.73 & AzTEC & \citet{Zeballos2018}\\
        MRC 0355-037 & H$z$RG(ELAN) & 2.15 & 108 & 120 & 6 & 0.78 & AzTEC & \citet{Zeballos2018}\\
        Spiderweb & H$z$RG(ELAN) & 2.16 & 263 & 90 & 12 & 0.70 & AzTEC & \citet{Zeballos2018}\\
        Slug & ELAN & 2.28 & 450 & -50 & 2 & 1.02 & SCUBA-2 & \citet{Nowotka2022}\\
        MAMMOTH-1 & ELAN & 2.32 & 440 & -77 & 10 & 0.5 & SCUBA-2 & \citet{AB2023}\\
        MRC 2104-242 & H$z$RG(ELAN) & 2.49 & 129  & 12 & 5 & 0.83 & AzTEC & \citet{Zeballos2018}\\
        PKS 0529-549 & H$z$RG(LAB) & 2.58 & 47 & 104 & 3 & 0.62 & AzTEC & \citet{Zeballos2018}\\
        \hline
    \end{tabular}
    \tablefoot{Column 1: source name; column 2: galaxy type; column 3: redshift; column 4: size of the Ly$\alpha$ nebula \citep{AB15}; column 5: major axis direction of the radio emission or Ly$\alpha$ nebula of H$z$RG or ELAN \citep{Zeballos2018,AB2023}; column 6: number of SMGs
    with de-boosted flux $S_{850\,{\rm \mu m}}\,{\gtrsim}\,5$\,mJy or $S_{1.1\rm mm}\,{\gtrsim}\,3.4$\,mJy
    within the central 4\arcmin; column 7: RMS level at the center of each map. For fields with fewer sources, this paucity might be partially caused by the lower sensitivity and completeness.}
\end{table*}

\subsection{Origin of the bright SMG overdensity} \label{bright}

Our overdensity analysis shows that only the number of SMGs with $S_{850\,{\rm \mu m}}\,{\gtrsim}\,5$\,mJy is significantly higher than the blank field counts.

If SMGs have the same origin, faint SMGs should also be overdense in the 4C\,23.56 field. The reason of no significant overdensity found for faint sources can be that numerous faint SMGs in the blank field make the number excess less obvious, or the blank field number counts derived from the shallower surveys, are not accurate at the faint end. This has been suggested by deep ALMA studies \citep[e.g.,][]{Stach2018,Gonzalez2021}, where the number counts are seen to turn flat at low fluxes; however, this is still under debate \citep[e.g.,][]{Fujimoto2023}. If we use the number counts from \citet{Stach2018} instead, the faint sources also show a number excess (see Fig.~\ref{fig:nc}).

On the other hand, the overdensity of bright SMGs can be caused by the source multiplicity. Previous studies have indicated that ${>}30\%$ bright SMGs would be split into multiple fainter SMGs with a high-resolution observation \citep[e.g.,][]{Hodge2013, Stach2018, Hayward2018, chen2022}.
Table~\ref{tab:SMA} shows the seven brightest SCUBA-2 sources with their corresponding IDs and flux densities from SMA data. The flux density measured from SMA is consistent with the SCUBA-2 measurement and none of these sources show obvious multiplicity.
It is thus unlikely that overdensity at the bright end is mainly caused by the source multiplicity.
As can be seen from Fig.~\ref{fig:ovdr}, these intrinsically bright SMGs tend to avoid the ``core'' region defined by the central radio galaxy,
which is consistent with the scenario that the bright SMG overdensity traces the extended star formation in the protocluster \citep{Chiang2017}.

\begin{table}[tp]
    \caption{Flux densities of the seven brightest SCUBA-2 sources.}
    \label{tab:SMA}
    \centering
    \begin{tabular}{cccc}
    \hline
    \hline
        ID & $S_{850\,{\rm \mu m}}$ & $S_{\rm d}$ & $S_{\rm SMA}$ \\
           & [mJy] & [mJy] & [mJy] \\
        \hline
        1 & 10.3$\pm$0.6 & 10.2$\pm$1.1 & 13.2$\pm$1.9  \\
        2 & 8.3$\pm$0.7 & 7.8$\pm$1.0 & ${<}\,10.0$ \\
        3 & 10.7$\pm$0.9 & 9.4$\pm$1.4 & 10.4$\pm$1.6 \\
        4 & 7.4$\pm$0.7 & 7.1$\pm$0.8 & 7.6$\pm$1.8 \\
        5 & 11.6$\pm$1.1 & 10.6$\pm$1.7 & 10.3$\pm$2.5 \\
        10 & 8.4$\pm$1.1 & 7.6$\pm$1.1 & 9.1$\pm$2.5 \\
        19 & 8.9$\pm$1.4 & 7.5$\pm$1.8 & 13.3$\pm$2.9 \\
        \hline
    \end{tabular}
     \tablefoot{Column 2: SCUBA-2 $850\,{\rm \mu m}$ fluxes; column 3: de-boosted SCUBA-2 fluxes; column 4: SMA 870+750~$\mu$m fluxes.
    The small discrepancies indicate that those SMGs are intrinsically bright.}
\end{table}

If multiplicity is not the reason of the bright SMG overdensity, the explanation can be that the protocluster 4C\,23.56 contains massive gas reservoirs, as the long-wavelength dust continuum can be a proxy of cold dust and gas \citep[e.g.,][]{Scoville2017,Scoville2022}.
Protoclusters are located in deep potential wells, and the cold gas can be efficiently supplied through the surrounding filamentary structures \citep[e.g.,][]{Tadaki2019,Daddi2021,Aoyama2022}. The molecular gas density in protoclusters is therefore higher than that in the field \citep[e.g.,][]{Lee2017,Jin2021,Polletta2022}. These gas-rich galaxies tend to show brighter dust continuum, which causes the number excess of bright SMGs.

\section{Conclusions} \label{s6}
In this paper we have presented the results from a pilot study of the JCMT/SCUBA-2 large program RAGERS.
We studied the overdensity and the spatial distribution of SMGs in the vicinity of H$z$RG 4C\,23.56 with SCUBA-2 $850\,{\rm \mu m}$ data.
Our main results are summarized as follows.
   \begin{enumerate}
      \item Our number counts suggest that the overdensity is pronounced for bright SMGs and that the faint-end counts are consistent with the blank field.
      \item The overall overdensity is more significant within the central 4\arcmin, which is the typical size of protoclusters at cosmic noon. However, the number counts in the outer region are also higher than those in the blank field.
      \item We find a positive correlation between flux and overdensity. Faint SMGs are concentrated within the central 2\arcmin, whereas bright SMGs show a significant excess at $3\arcmin\,{\leq}\,r\,{<}\,4\arcmin$.
      \item The angular overdensity analysis indicates that the overdensity is more significant along the direction perpendicular to the radio jet. Both bright and faint SMGs show an angular preference in both inner and outer regions. In the jet region, the SMG distribution shows less dependence on the flux density and distance.
      \item We used MMP$z$ to estimate the photo-$z$ based on the available FIR/submillimeter data. More bright SMGs ($S_{850\,{\rm \mu m}}\,{\geq}\,5$\,mJy) are close to the redshift of the H$z$RG. The anisotropic distribution still holds when only 29 promising candidates remain.
      \item High-resolution SMA observations ($\sim$\,2\arcsec) show that the seven brightest SMGs are intrinsically bright and do not break up into multiple sources. This suggests that the overdensity of bright SMGs is not due to source multiplicity.
      \item Compared to other fields at similar redshifts, the SMGs are less concentrated in the inner region and show a higher anisotropic level in the 4C\,23.56 field, which suggests that the 4C\,23.56 protocluster might be at a different evolutionary stage. But a firm conclusion cannot be made due to the relatively small sample size and FoV as well as the lack of spectroscopic redshifts. A larger sample with spectroscopic information is needed for further exploration.
   \end{enumerate}

It is likely that the spatial distribution of SMGs reveals the interplay between the radio AGN and the surrounding obscured star formation activity. This case study, however, does not allow us to draw any general conclusion for protoclusters traced by H$z$RGs.
The origins of these possible correlations need to be investigated using a larger sample. Next-generation cameras, such as the W-band polarimetry Imager using Kinetic Inductance Detectors (WIKID) and TolTEC \citep{toltec}, and single-dish telescopes, including Large-Sized Telescope \citep[LST,][]{lst1,lst2}, Atacama Large Aperture Submillimeter Telescope \citep[AtLAST,][]{atlast, Ramasawmy2022, Mroczkowski2023}, and Fred Young Submillimeter Telescope \citep[FYST or CCAT-prime,][]{ccat}, will be ideal for follow-up studies.

\section*{Data availability}
The SCUBA-2 $850\,{\rm \mu m}$ data and scripts used for the de-boosting process are available on github: \url{https://github.com/dazhiUBC/SCUBA2_MF}.

\begin{acknowledgements}
We thank the referee for a very helpful report in reviewing this manuscript. We are grateful to Dr.~Caitlin Casey for the useful discussions and her help with the MMP$z$ code. We thank Dr. Katherine Blundell for providing us with the VLA 5GHz image in B configuration (AC234) and Dr.~Shuowen Jin for his suggestions about SPIRE data analysis. We thank George Wang for useful comments and discussions. The Cosmic Dawn Center (DAWN) is funded by the Danish National Research Foundation under grant No.~140. TRG and ML are grateful for support from the Carlsberg Foundation via grant No.~CF20-0534. B.G. acknowledges support from the Carlsberg Foundation Research Grant CF20-0644 ‘Physical pRoperties of the InterStellar Medium in Luminous Infrared Galaxies at High redshifT: PRISMLIGHT’ LCH was supported by the National Science Foundation of China (11721303, 11991052, 12011540375, 12233001) and the China Manned Space Project (CMS-CSST-2021-A04, CMS-CSST-2021-A06). C.-C.C. and Y.-J.W acknowledged support from the National Science and Technology Council of Taiwan (NSTC 109-2112-M-001-016-MY3 and 111-2112M-001-045-MY3), as well as Academia Sinica through the Career Development Award (AS-CDA-112-M02). X.-J.J. acknowledges support from the National Science Foundation of China (12373026). KK acknowledges the support by JSPS KAKENHI Grant Numbers JP17H06130, JP22H04939, and JP23K20035. This work has been supported by the French government, through the UCA\textsuperscript{J.E.D.I.} Investments in the Future project managed by the National Research Agency (ANR) with the reference number ANR-15-IDEX-01. The\textit{ James Clerk} Maxwell Telescope is operated by the East Asian Observatory on behalf of The National Astronomical Observatory of Japan; Academia Sinica Institute of Astronomy and Astrophysics; the Korea Astronomy and Space Science Institute; the National Astronomical Research Institute of Thailand; Center for Astronomical Mega-Science (as well as the National Key R\&D Program of China with No. 2017YFA0402700). Additional funding support is provided by the Science and Technology Facilities Council of the United Kingdom and participating universities and organizations in the United Kingdom and Canada.
Additional funds for the construction of SCUBA-2 were provided by the Canada Foundation for Innovation. The \textit{James Clerk} Maxwell Telescope has historically been operated by the Joint Astronomy Centre on behalf of the Science and Technology Facilities Council of the United Kingdom, the National Research Council of Canada and the Netherlands Organisation for Scientific Research. The authors wish to recognize and acknowledge the very significant cultural role and reverence that the summit of Maunakea has always had within the indigenous Hawaiian community. We are most fortunate to have the opportunity to conduct observations from this mountain. The Starlink software \citep{Currie2014} is currently supported by the East Asian Observatory. This research has made use of data obtained from the Chandra Data Archive and the Chandra Source Catalog, and software provided by the Chandra X-ray Center (CXC) in the application packages CIAO and Sherpa. 
Herschel is an ESA space observatory with science instruments provided by European-led Principal Investigator consortia and with important participation from NASA. 
The ASTE project driven by Nobeyama Radio Observatory (NRO), a branch of National Astronomical Observatory of Japan (NAOJ), in collaboration with University of Chile, and Japanese institutes including University of Tokyo, Nagoya University, Osaka Prefecture University, Ibaraki University, and Hokkaido University. Observations with ASTE were in part carried out remotely from Japan by using NTT's GEMnet2 and its partnet R\&E (Research and Education) networks, which are based on AccessNova collaboration of University of Chile, NTT Laboratories, and NAOJ. The Submillimeter Array is a joint project between the Smithsonian Astrophysical Observatory and the Academia Sinica Institute of Astronomy and Astrophysics and is funded by the Smithsonian Institution and the Academia Sinica. 
The Green Bank Observatory is a facility of the National Science Foundation operated under cooperative agreement by Associated Universities, Inc. 
This research made use of Astropy, a community-developed core Python package for Astronomy \citep{astropy1,Astropy2,Astropy3}, SciPy \citep{scipy}, NumPy \citep{numpy} and Matplotlib, a Python library for publication quality graphics \citep{plt}.
\end{acknowledgements}

\bibliographystyle{aa}
\bibliography{aanda.bib}
\begin{appendix}
\onecolumn
\section{\texorpdfstring{$850\,{\rm \mu m}$}{850} source catalog} \label{catalog}
Here we show $850\,{\rm \mu m}$ the source catalog (S/N\,${\geq}\,4\sigma$) with fluxes and uncertainties before and after de-boosting (Table.\,\ref{850table}).
\FloatBarrier
\begin{table*}[!h]
    \caption{Positions and fluxes of the $850\,{\rm \mu m}$-selected SMGs. The uncertainty is the combined total 1$\sigma$ uncertainty.}
    \label{850table} 
    \centering
    \begin{tabular}{lcccccccc}
    \hline
      ID &        RA &       Dec &  flux &  RMS  &   S/N & de-boosting factor & flux$_{\rm d}$  & uncertainty \\
      &[deg]&[deg]&[mJy]&[mJy]&&&[mJy] & [mJy] \\
      \hline
      1 & 21.12211 & 23.5147 &  10.3 &  0.6 & 17.8 &  0.99&   10.2 &          1.1 \\
      2 & 21.12041 & 23.5652 &   8.3 &  0.7 & 12.4 &  0.94&    7.8 &          1.0 \\
      3 & 21.11508 & 23.5352 &  10.7 &  0.9 & 11.9 &  0.88&    9.4 &          1.4 \\
      4 & 21.11900 & 23.5736 &   7.4 &  0.7 & 10.9 &  0.96&    7.1 &          0.8 \\
      5 & 21.12728 & 23.4758 &  11.6 &  1.1 & 10.7 &  0.91&   10.6 &          1.7 \\
      6 & 21.12393 & 23.5647 &   6.1 &  0.7 &  9.0 &  0.95&    5.8 &          1.0 \\
      7 & 21.11908 & 23.5586 &   5.8 &  0.7 &  8.5 &  0.97&    5.6 &          1.0 \\
      8 & 21.11928 & 23.4802 &   6.4 &  0.8 &  8.3 &  0.94&    6.0 &          1.1 \\
      9 & 21.11936 & 23.5902 &   6.1 &  0.8 &  8.0 &  0.95&    5.8 &          1.1 \\
     10 & 21.11690 & 23.4686 &   8.4 &  1.1 &  7.6 &  0.90&    7.6 &          1.1 \\
     11 & 21.12437 & 23.4852 &   5.6 &  0.8 &  7.3 &  0.93&    5.2 &          1.1 \\
     12 & 21.12692 & 23.5658 &   6.3 &  0.9 &  7.2 &  0.94&    5.9 &          1.1 \\
     13 & 21.12498 & 23.5208 &   5.7 &  0.8 &  7.2 &  0.91&    5.2 &          1.2 \\
     14 & 21.11650 & 23.5080 &   5.8 &  0.8 &  7.1 &  0.95&    5.5 &          1.0 \\
     15 & 21.12009 & 23.5036 &   4.5 &  0.7 &  6.7 &  0.87&    3.9 &          1.1 \\
     16 & 21.12272 & 23.5558 &   3.9 &  0.6 &  6.6 &  0.85&    3.3 &          1.0 \\
     17 & 21.12134 & 23.5319 &   3.6 &  0.6 &  6.5 &  0.89&    3.2 &          0.9 \\
     18 & 21.12660 & 23.5419 &   5.7 &  0.9 &  6.4 &  0.89&    5.1 &          1.1 \\
     19 & 21.11468 & 23.5802 &   8.9 &  1.4 &  6.3 &  0.84&    7.5 &          1.8 \\
     20 & 21.11609 & 23.5319 &   4.8 &  0.8 &  5.9 &  0.83&    4.0 &          1.1 \\
     21 & 21.12138 & 23.5114 &   3.5 &  0.6 &  5.8 &  0.83&    2.9 &          0.8 \\
     22 & 21.12025 & 23.5941 &   4.8 &  0.8 &  5.8 &  0.83&    4.0 &          1.2 \\
     23 & 21.12296 & 23.5297 &   3.3 &  0.6 &  5.6 &  0.79&    2.6 &          0.9 \\
     24 & 21.11965 & 23.5686 &   3.7 &  0.7 &  5.4 &  0.81&    3.0 &          1.0 \\
     25 & 21.12518 & 23.4836 &   4.2 &  0.8 &  5.3 &  0.83&    3.5 &          0.9 \\
     26 & 21.11702 & 23.5536 &   3.9 &  0.8 &  5.2 &  0.79&    3.1 &          1.1 \\
     27 & 21.12247 & 23.5008 &   3.5 &  0.7 &  5.2 &  0.80&    2.8 &          0.9 \\
     28 & 21.12199 & 23.5158 &   2.9 &  0.6 &  5.1 &  0.83&    2.4 &          0.9 \\
     29 & 21.12869 & 23.4524 &  11.3 &  2.2 &  5.0 &  0.73&    8.2 &          2.2 \\
     30 & 21.12256 & 23.5380 &   2.7 &  0.6 &  4.8 &  0.81&    2.2 &          0.8 \\
     31 & 21.12817 & 23.5702 &   4.8 &  1.0 &  4.8 &  0.75&    3.6 &          1.4 \\
     32 & 21.11977 & 23.4464 &   5.1 &  1.1 &  4.8 &  0.80&    4.1 &          1.4 \\
     33 & 21.12163 & 23.5208 &   2.6 &  0.6 &  4.6 &  0.81&    2.1 &          0.8 \\
     34 & 21.12865 & 23.4602 &   9.5 &  2.1 &  4.6 &  0.66&    6.3 &          2.2 \\
     35 & 21.11981 & 23.5508 &   3.0 &  0.7 &  4.6 &  0.80&    2.4 &          0.9 \\
     36 & 21.12256 & 23.5219 &   2.6 &  0.6 &  4.5 &  0.81&    2.1 &          0.8 \\
     37 & 21.12045 & 23.5675 &   3.1 &  0.7 &  4.5 &  0.71&    2.2 &          1.0 \\
     38 & 21.12078 & 23.5214 &   2.6 &  0.6 &  4.4 &  0.69&    1.8 &          0.9 \\
     39 & 21.12377 & 23.4736 &   3.8 &  0.9 &  4.3 &  0.74&    2.8 &          1.2 \\
     40 & 21.12663 & 23.4358 &   8.5 &  2.0 &  4.3 &  0.58&    4.9 &          2.6 \\
     41 & 21.12264 & 23.5297 &   2.5 &  0.6 &  4.3 &  0.68&    1.7 &          0.9 \\
     42 & 21.12878 & 23.5469 &   4.2 &  1.0 &  4.2 &  0.64&    2.7 &          1.4 \\
     43 & 21.12074 & 23.3886 &   9.8 &  2.3 &  4.2 &  0.73&    7.2 &          2.2 \\
     44 & 21.12704 & 23.6024 &   5.2 &  1.2 &  4.2 &  0.71&    3.7 &          1.4 \\
     45 & 21.12377 & 23.5269 &   2.7 &  0.6 &  4.2 &  0.78&    2.1 &          0.8 \\
     46 & 21.12821 & 23.5547 &   3.9 &  0.9 &  4.1 &  0.69&    2.7 &          1.2 \\
     47 & 21.12033 & 23.5630 &   2.7 &  0.7 &  4.1 &  0.74&    2.0 &          0.8 \\
     48 & 21.12954 & 23.4946 &   6.3 &  1.5 &  4.1 &  0.62&    3.9 &          2.0 \\
     49 & 21.12078 & 23.5302 &   2.3 &  0.6 &  4.0 &  0.70&    1.6 &          0.8 \\
     50 & 21.11318 & 23.5624 &   7.4 &  1.9 &  4.0 &  0.58&    4.3 &          2.7 \\
     \hline
     \end{tabular}
\end{table*}
\section{Photometry and photo-z}
We include the corresponding FIR catalog with SPIRE, AzTEC, and MUSTANG-2 photometry (Table.\,\ref{multi-z}). The photo-$z$ derived from MMP$z$ is also provided with upper- and lower uncertainties.
\def\arraystretch{1.185}
{\tiny 
\begin{table*}
    \caption{De-boosted photometry and redshift estimation from MMP$z$.}
    \label{multi-z}
    \centering
    \begin{tabular}{lccccccc}
    \hline
     ID &  $S_{250\,\rm \mu m}$[mJy] &   $S_{350\,\rm \mu m}$[mJy] &    $S_{500\,\rm \mu m}$[mJy] &  $S_{850\,\rm \mu m}$[mJy] &  $S_{1.1\,\rm mm}$[mJy] &  $S_{3.3\,\rm mm}$[mJy] & z \\
     \hline
    1 & 19.5$\pm$5.8 & 44.9$\pm$6.3 & 48.5$\pm$6.8 & 10.2$\pm$0.8 & 10.2$\pm$2.1 & 0.164$\pm$0.044 & 2.98$^{+1.59}_{-0.96}$ \\
    2 & 25.5$\pm$5.8 & 40.0$\pm$6.3 & 36.3$\pm$6.8 &  7.8$\pm$0.8 & 5.4$\pm$2.1 & 0.058$\pm$0.056 & 2.48$^{+0.66}_{-0.6}$ \\
    3 & 4.1$\pm$2.6 & 1.5$\pm$2.4 & 2.0$\pm$3.1 &  9.4$\pm$1.0 & 5.5$\pm$2.1 & 0.379$\pm$0.165 & 7.97$^{+1.93}_{-1.75}$ \\
    4 & 22.3$\pm$5.8 & 29.2$\pm$6.3 & 30.8$\pm$6.8 &  7.1$\pm$0.8 & 4.0$\pm$2.1 & 0.186$\pm$0.063 & 2.67$^{+1.03}_{-0.79}$ \\
    5 & 31.9$\pm$5.8 & 8.4$\pm$6.3 & 16.9$\pm$6.8 & 10.6$\pm$1.2 & 5.7$\pm$2.1 & nan & 3.42$^{+3.71}_{-1.44}$ \\
    6 & 12.7$\pm$5.8 & 23.6$\pm$6.3 & 16.9$\pm$6.8 &  5.8$\pm$0.8 & 5.3$\pm$2.1 & 0.062$\pm$0.080 & 2.83$^{+0.82}_{-0.71}$ \\
    7 & 9.5$\pm$5.8 & 17.7$\pm$6.3 & 16.3$\pm$6.8 & 5.6$\pm$0.8 & 3.6$\pm$2.1 & 0.0$\pm$0.056 & 2.88$^{+0.87}_{-0.65}$ \\
    8 & 36.3$\pm$5.8 & 40.6$\pm$6.3 & 32.9$\pm$6.8 &  6.0$\pm$0.8 & 4.0$\pm$2.1 & 0.17$\pm$0.076 & 1.98$^{+0.65}_{-0.6}$ \\
    9 & 24.9$\pm$5.8 & 35.9$\pm$6.3 & 31.8$\pm$6.8 &  5.8$\pm$0.8 & 3.6$\pm$2.1 & 0.007$\pm$0.098 & 2.12$^{+0.78}_{-0.6}$ \\
    10 & 29.7$\pm$5.8 & 42.9$\pm$6.3 & 36.8$\pm$6.8 &  7.6$\pm$1.2 & 3.7$\pm$2.1 & 0.162$\pm$0.167 & 2.38$^{+0.63}_{-0.61}$ \\
    11 & 6.2$\pm$5.8 & 4.9$\pm$2.1 & 8.8$\pm$6.8 &  5.2$\pm$0.9 & 3.6$\pm$2.1 & 0.045$\pm$0.120 & 4.67$^{+1.15}_{-0.9}$ \\
    12 & 19.7$\pm$5.8 & 13.8$\pm$6.3 & 7.2$\pm$6.8 &  5.9$\pm$1.0 & 4.2$\pm$2.1 & nan & 2.92$^{+0.63}_{-0.59}$ \\
    13 & 9.5$\pm$5.8 & 17.6$\pm$6.3 & 11.1$\pm$6.8 &  5.2$\pm$0.9 & 3.6$\pm$2.1 & 0.158$\pm$0.089 & 3.12$^{+0.92}_{-0.7}$ \\
    14 & 15.4$\pm$5.8 & 32.3$\pm$6.3 & 41.0$\pm$6.8 &  5.5$\pm$0.9 & 3.9$\pm$2.1 & 0.166$\pm$0.097 & 2.52$^{+3.15}_{-1.59}$ \\
    15 & 6.8$\pm$5.8 & 4.0$\pm$2.3 & 2.8$\pm$2.8 &  3.9$\pm$0.8 & 1.9$\pm$2.1 & 0.075$\pm$0.048 & 4.78$^{+1.52}_{-1.3}$ \\
    16 & 46.5$\pm$5.8 & 23.7$\pm$6.3 & 5.8$\pm$6.8 &  3.3$\pm$0.8 & 1.0$\pm$0.6 & 0.036$\pm$0.057 & 0.82$^{+0.39}_{-0.54}$ \\
    17 & 9.7$\pm$5.8 & 22.5$\pm$6.3 & 8.1$\pm$6.8 &  3.2$\pm$0.8 & 3.9$\pm$2.1 & 0.055$\pm$0.045 & 2.27$^{+2.17}_{-1.2}$ \\
    18 & 5.6$\pm$5.8 & 13.5$\pm$6.3 & 8.1$\pm$6.8 &  5.1$\pm$1.0 & 3.6$\pm$2.1 & 0.228$\pm$0.178 & 3.73$^{+1.14}_{-0.9}$ \\
    19 & 19.9$\pm$5.8 & 29.1$\pm$6.3 & 25.4$\pm$6.8 &  7.5$\pm$1.6 & 6.3$\pm$2.1 & nan & 2.88$^{+0.68}_{-0.65}$ \\
    20 & 36.1$\pm$5.8 & 36.1$\pm$6.3 & 17.1$\pm$6.8 &  4.0$\pm$0.9 & 1.8$\pm$2.1 & 0.199$\pm$0.099 & 1.52$^{+0.57}_{-0.7}$ \\
    21 & 0.0$\pm$2.3 & 0.0$\pm$2.4 & 0.0$\pm$2.8 &  2.9$\pm$0.8 & 2.8$\pm$2.1 & 0.018$\pm$0.047 & 5.42$^{+2.43}_{-1.9}$ \\
    22 & 0.0$\pm$2.5 & 6.2$\pm$6.3 & 7.4$\pm$6.8 &  4.0$\pm$0.9 & 2.5$\pm$2.1 & 0.004$\pm$0.102 & 4.67$^{+1.58}_{-1.15}$ \\
    23 & 4.3$\pm$2.3 & 6.3$\pm$6.3 & 14.1$\pm$6.8 &  2.6$\pm$0.8 & 2.3$\pm$2.1 & 0.052$\pm$0.050 & 3.12$^{+1.98}_{-1.14}$ \\
    24 & 2.9$\pm$2.4 & 9.0$\pm$6.3 & 0.0$\pm$2.8 &  3.0$\pm$0.8 & 0.9$\pm$0.6 & 0.0$\pm$0.062 & 3.58$^{+4.32}_{-1.85}$ \\
    25 & 29.2$\pm$5.8 & 40.1$\pm$6.3 & 22.9$\pm$6.8 &  3.5$\pm$0.9 & 2.7$\pm$2.1 & 0.08$\pm$0.171 & 1.57$^{+0.82}_{-0.79}$ \\
    26 & 5.2$\pm$5.8 & 4.8$\pm$2.8 & 6.4$\pm$6.8 &  3.1$\pm$0.8 & 0.0$\pm$0.7 & 0.131$\pm$0.081 & 3.08$^{+4.87}_{-2.26}$ \\
    27 & 0.0$\pm$2.2 & 0.0$\pm$2.3 & 0.0$\pm$2.6 &  2.8$\pm$0.8 & 0.5$\pm$0.6 & 0.033$\pm$0.057 & 5.33$^{+3.95}_{-2.31}$ \\
    28 & 5.5$\pm$5.8 & 36.7$\pm$6.3 & 40.1$\pm$6.8 &  2.4$\pm$0.8 & 8.5$\pm$2.1 & 0.087$\pm$0.048 & 2.42$^{+5.17}_{-2.39}$ \\
    29 & 883.5$\pm$8.6 & 363.1$\pm$6.3 & 122.1$\pm$6.8 &  8.2$\pm$2.5 & 6.6$\pm$2.1 & nan & 0.17$^{+0.13}_{-0.14}$ \\
    30 & 26.4$\pm$5.8 & 22.4$\pm$6.3 & 7.8$\pm$6.8 &  2.2$\pm$0.8 & 0.6$\pm$0.6 & 0.088$\pm$0.047 & 1.07$^{+0.62}_{-0.84}$ \\
    31 & 20.7$\pm$5.8 & 23.4$\pm$6.3 & 19.1$\pm$6.8 &  3.6$\pm$1.1 & 3.1$\pm$2.1 & nan & 1.93$^{+0.58}_{-0.55}$ \\
    32 & 9.9$\pm$5.8 & 11.6$\pm$6.3 & 11.3$\pm$6.8 &  4.1$\pm$1.2 & 1.9$\pm$2.1 & 0.457$\pm$0.194 & 3.02$^{+5.01}_{-1.45}$ \\
    33 & 15.6$\pm$5.8 & 36.5$\pm$6.3 & 32.5$\pm$6.8 &  2.2$\pm$0.8 & 3.3$\pm$2.1 & 0.012$\pm$0.046 & 1.73$^{+4.46}_{-1.7}$ \\
    34 & 7.2$\pm$5.8 & 6.0$\pm$6.3 & 34.4$\pm$6.8 &  6.3$\pm$2.3 & 6.0$\pm$2.1 & nan & 4.22$^{+4.67}_{-1.99}$ \\
    35 & 0.0$\pm$2.3 & 11.3$\pm$6.3 & 9.8$\pm$6.8 &  2.4$\pm$0.8 & 1.5$\pm$0.6 & 0.082$\pm$0.047 & 4.42$^{+4.02}_{-1.59}$ \\
    36 & 0.0$\pm$2.5 & 0.0$\pm$2.3 & 29.9$\pm$6.8 &  2.2$\pm$0.8 & 2.0$\pm$2.1 & 0.057$\pm$0.050 & 5.22$^{+5.28}_{-2.24}$ \\
    37 & 10.6$\pm$5.8 & 13.2$\pm$6.3 & 28.3$\pm$6.8 &  2.2$\pm$0.8 & 4.0$\pm$2.1 & 0.004$\pm$0.058 & 2.12$^{+4.44}_{-2.09}$ \\
    38 & 0.0$\pm$2.4 & 22.5$\pm$6.3 & 26.2$\pm$6.8 &  1.8$\pm$0.8 & 2.8$\pm$2.1 & 0.07$\pm$0.045 & 3.83$^{+5.93}_{-1.71}$ \\
    39 & 0.2$\pm$2.4 & 0.0$\pm$2.4 & 0.0$\pm$2.9 &  2.8$\pm$1.0 & 0.5$\pm$0.6 & 0.114$\pm$0.125 & 5.47$^{+4.98}_{-1.89}$ \\
    40 & 14.5$\pm$5.8 & 8.8$\pm$6.3 & 7.7$\pm$6.8 &  4.9$\pm$2.2 & 1.7$\pm$2.1 & nan & 2.52$^{+0.93}_{-0.9}$ \\
    41 & 13.6$\pm$5.8 & 16.9$\pm$6.3 & 18.3$\pm$6.8 &  1.7$\pm$0.8 & 2.0$\pm$2.1 & 0.0$\pm$0.048 & 1.57$^{+1.5}_{-1.54}$ \\
    42 & 1.4$\pm$3.0 & 6.4$\pm$6.3 & 9.9$\pm$6.8 &  2.7$\pm$1.1 & 2.1$\pm$2.1 & nan & 3.83$^{+2.93}_{-1.35}$ \\
    43 & 3.0$\pm$2.6 & 4.6$\pm$3.0 & 13.9$\pm$6.8 &  7.2$\pm$2.6 & 3.6$\pm$2.1 & nan & 5.12$^{+1.59}_{-1.24}$ \\
    44 & 0.0$\pm$2.5 & 0.0$\pm$2.6 & 0.0$\pm$3.1 &  3.7$\pm$1.4 & 0.0$\pm$0.8 & nan & 5.53$^{+5.17}_{-2.36}$ \\
    45 & 11.6$\pm$5.8 & 4.9$\pm$2.1 & 8.1$\pm$6.8 &  2.1$\pm$0.8 & 0.9$\pm$0.6 & 0.076$\pm$0.062 & 2.67$^{+1.58}_{-1.6}$ \\
    46 & 13.2$\pm$5.8 & 14.7$\pm$6.3 & 9.7$\pm$6.8 &  2.7$\pm$1.0 & 1.0$\pm$0.7 & nan & 1.88$^{+0.39}_{-0.4}$ \\
    47 & 20.3$\pm$5.8 & 30.7$\pm$6.3 & 29.3$\pm$6.8 &  2.0$\pm$0.8 & 4.2$\pm$2.1 & 0.074$\pm$0.055 & 1.57$^{+3.94}_{-1.54}$ \\
    48 & 0.0$\pm$2.5 & 0.0$\pm$2.6 & 1.5$\pm$2.9 &  3.9$\pm$1.7 & 1.5$\pm$2.1 & nan & 7.72$^{+2.99}_{-1.8}$ \\
    49 & 8.8$\pm$5.8 & 8.3$\pm$6.3 & 8.6$\pm$6.8 &  1.6$\pm$0.8 & 2.7$\pm$2.1 & 0.146$\pm$0.044 & 9.12$^{+2.88}_{-3.7}$ \\
    50 & 22.6$\pm$5.8 & 26.6$\pm$6.3 & 19.4$\pm$6.8 &  4.3$\pm$2.0 & 1.2$\pm$1.0 & nan & 1.88$^{+0.52}_{-0.65}$ \\
    \hline
    \end{tabular}
    \tablefoot{Due to the smaller coverage of MUSTANG-2 data, a few SMGs are outside the edge of its FoV, which are denoted as ``nan'' in the table.}
\end{table*}
}
\twocolumn
\section{Unusually faint \texorpdfstring{$3\,{\rm mm}$}{3mm} fluxes: Overestimation or contamination?} \label{3mm}
\begin{figure*}
    \centering
    \includegraphics[width=0.96\textwidth]{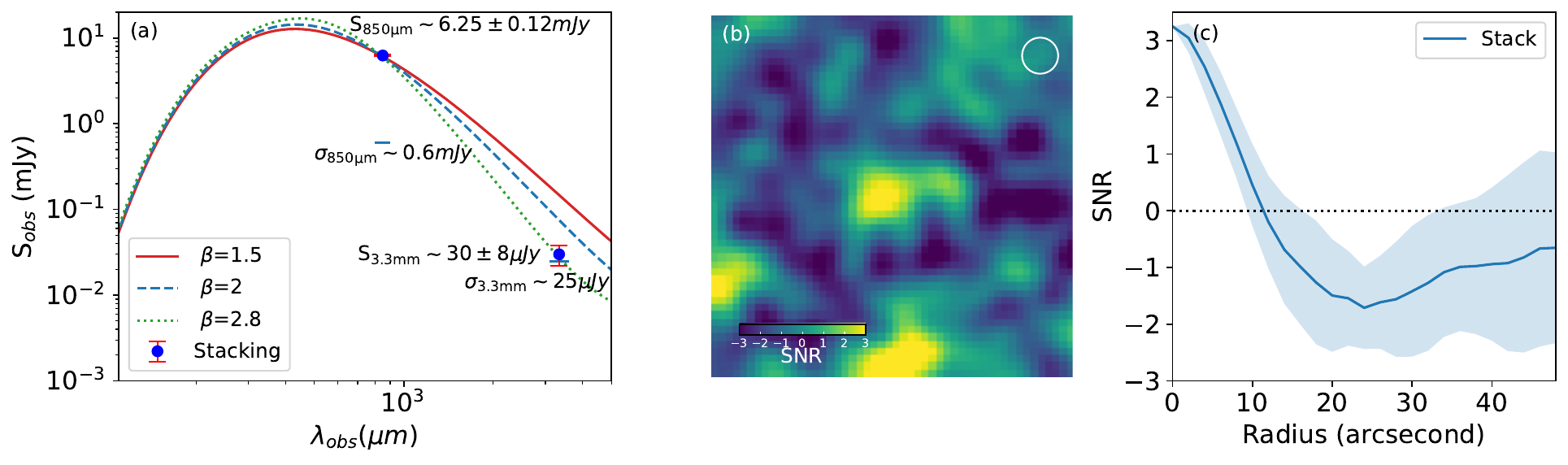} 
    \caption{Evidence of the unusually faint $3\,{\rm mm}$ continuum of SMGs in the 4C\,23.56 field. (a): SEDs for SMGs ($T_{\rm d}\,{=}\,35K$, $\beta\,{=}\,1.5$, 2, 2.8, $z\,{=}\,2.48$) with flux scale to the stacked $850\,{\rm \mu m}$ flux. The observed $3\,{\rm mm}$ flux can be only reproduced when $\beta\,{\sim}\,2.8$. (b): S/N map of the stacked image for the 38 SMGs with a distance between 1\arcmin\, and 4\arcmin\, from the H$z$RG. A negative ring is shown around the central source. The white circle denotes the MUSTANG-2 beam size. (c): Azimuthally averaged S/N profile of the stacked source with the corresponding standard deviations in each annulus. }
    \label{fig:stack}
\end{figure*}

To further characterize the long-wavelength SEDs and robust dust properties for $850\,{\rm \mu m}$-selected sources, we conducted MUSTANG-2 observation.
The sensitivity of MUSTANG-2 observation at $3\,{\rm mm}$ (${\sim}\,21\,{\rm \mu Jy}$) would allow us to detect most SMGs with infrared luminosity above $2\times10^{12}L_\odot$.
However, even when utilizing a 2$\sigma$ detection threshold, the actual number of detections falls short of expectations. This discrepancy might align with the previously reported steep dust SEDs \citep[e.g.,][]{dacunha2021,Cooper2022,Jin2022}, suggesting a common overestimation of the dust continuum at long wavelengths. We conducted a stacking analysis for all 38 $850\,{\rm \mu m}$-selected sources located between 1\arcmin\ and 4\arcmin\ from the H$z$RG. The stacked fluxes at $850\,{\rm \mu m}$ and $3\,{\rm mm}$ are ${\sim}\,6.25\pm0.12,{\rm mJy}$ and ${\sim}\,30\pm8,{\rm \mu Jy}$, respectively. Assuming $z\,{=}\,2.48$, a dust temperature of $T_{\rm d}\,{=}\,35$K, and a critical frequency $\nu_c\,{=}\,1500$ GHz, the emissivity spectral index is estimated to be ${\sim}\,2.8$ (see Fig.~\ref{fig:stack}a). Because we do not expect many sources at $z\,{>}\,3$, the impact from cosmic microwave background should be small. This suggests that the steep Rayleigh-Jeans slope alone may not explain the low $3\,{\rm mm}$ flux with current models \citep[e.g.,][]{kohler2015}.

Alternatively, this discrepancy may be attributed to the SZ effect, which could lead to a significant underestimation of the intrinsic dust continuum at $3\,{\rm mm}$. Some compact negative signals (${<}\,{-}\,4\sigma$) show counterparts in our ancillary data, which could be potentially caused by the SZ effect from foreground sources at low redshifts and contaminate our result. Additionally, several expected SMGs exhibited negative $3\,{\rm mm}$ signals at their positions. We cannot rule out the possibility of a hot circumgalactic medium surrounding protocluster members, potentially causing the SZ decrement. Recent observations suggest that powerful AGNs could efficiently heat the circumgalactic medium and result in SZ decrements at high redshifts \citep{Hatch2014,Crichton2017,Hall2019,lacy2019,DiM2023,Dong2023,Meinke2023}. The simulation shows that the AGN heating is able to reproduce SZ decrements of ${\sim}\,{-}0.1$mJy and ${\sim}\,{-}0.5$mJy on small ($\sim$\,10\arcsec) and large ($\sim$\,40\arcsec) scales, respectively \citep{Henden2018,Brownson2019,Bennett2023}.

Given that the MUSTANG-2 resolution is limited to ${\sim}\,10\arcsec$, only the extended component is detectable. The stacked image reveals a negative ring around the source, down to $-3\sigma$ (Fig.~\ref{fig:stack}b), which is also observed in the radial profile (Fig.~\ref{fig:stack}c).
To ensure it is not caused by any introduced artifacts, we carefully checked if the decrement is due to the data reduction, filtering, or smoothing process. The power spectrum of the noise map is mostly flat, which suggests that the data product is dominated by the white noise that is uniform in different scales. We measured the PSF shape of a bright point source from another observation (S/N$\,>$\,400), the data of which  were reduced and filtered in a similar manner. The reduction and the filtering step cause a decrement only up to ${\sim}\,1\%$. We also repeated the stacking and smoothing process on random locations 5,000 times. Only two simulated maps has been found with similar radial profile. Therefore, the scenario of the artificial decrement is less preferred, suggesting the potential existence of extended SZ signals or the atmospheric contamination.
Thus, the observed faint $3\,{\rm mm}$ flux compared to expectations may indicate either an overestimation of the dust continuum at the long wavelength or the presence of the SZ effect in the H$z$RG environment. Further multiband long-wavelength observations are essential to distinguish between these possibilities and refine our understanding of the dust properties in this high-redshift environment.
\begin{figure}[!h]
    \centering
    \includegraphics[width=0.48\textwidth]{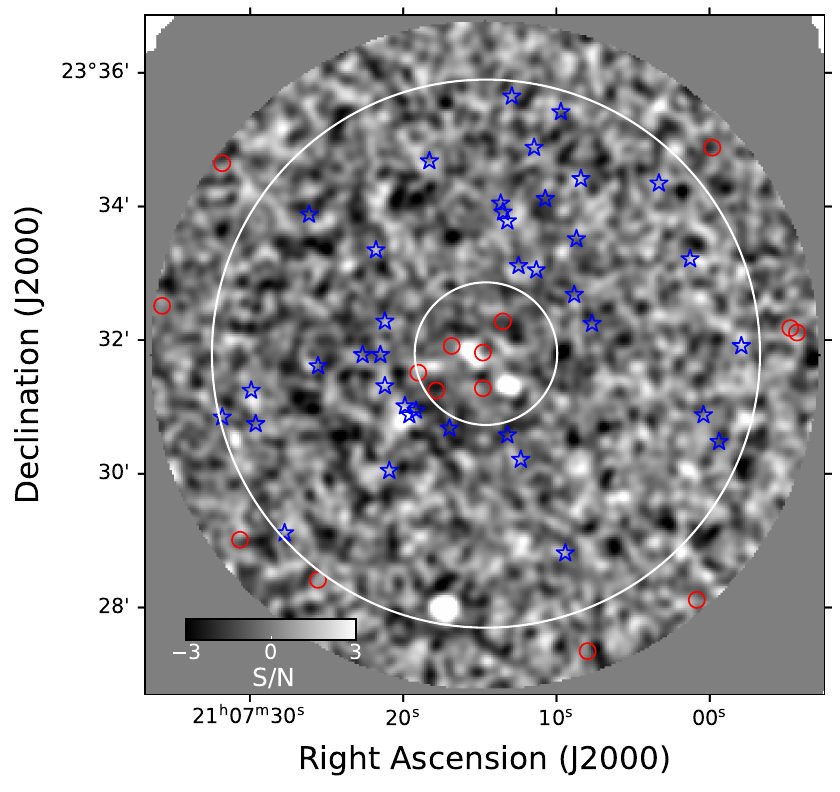}
    \caption{MUSTANG-2 S/N map at $3\,{\rm mm}$. 56 SMGs are covered. The two white circles are the inner and outer boundaries of the selected region for the stacking analysis, where 38 SMGs are included as blue stars. The rest are indicated as red circles. The brightest source is a low-z blazar, which is not associated with the protocluster.}
    \label{fig:m2}
\end{figure}
\end{appendix}
\end{CJK*}
\end{document}